\documentclass[aps,prd,reprint,floatfix,superscriptaddress,lengthcheck,sort&compress,letterpaper,nofootinbib]{revtex4-2}

\usepackage{amssymb}
\usepackage[intlimits]{amsmath}
\usepackage{amstext} 
\usepackage{amsfonts}
\usepackage{dsfont}
\usepackage{float}
\usepackage{slashed}
\usepackage{subfigure}
\usepackage{natbib}
\usepackage{multirow}
\usepackage{verbatim}
\usepackage{nicefrac}
\usepackage{dcolumn}
\usepackage{units}
\usepackage{bm}
\usepackage{wasysym}
\usepackage{array}
\usepackage{graphicx}
\usepackage[usenames]{color}
\usepackage[colorlinks=true,linkcolor=blue,citecolor=blue,urlcolor=blue]{hyperref}
\usepackage{tabularx}
\usepackage{makecell}

\newcommand{\conjg}[1]{\ensuremath{\hspace{1pt}\overline{\hspace{-1pt}#1\hspace{-1pt}}}\hspace{1pt}}

\def\mD{\ensuremath{\mathcal{D}}}

\def\mM{\ensuremath{\mathcal{M}}}
\def\mN{\ensuremath{\mathcal{N}}}
\def\mA{\ensuremath{\mathcal{A}}}
\def\mK{\ensuremath{\mathcal{K}}}
\def\mI{\ensuremath{\mathcal{I}}}

\def\PiT{\ensuremath{\widetilde\Pi}}
\def\p{\partial}
\newcommand{\vect}[1]{{\mbox{\boldmath $#1$}}}

      \definecolor{violet}{RGB}{111,0,255}
      \definecolor{webgreen}{rgb}{0,0.75,0}
      \definecolor{webred}{rgb}{0.75,0,0}
      \definecolor{webblue}{rgb}{0,0,0.75}
      \definecolor{darkblue}{rgb}{0,0,0.6}
      \definecolor{darkgreen}{rgb}{0,0.5,0.5}
      \definecolor{darkpurple}{rgb}{0.5,0,0.5}
      \definecolor{darkorange}{rgb}{1,0.5,0}
      \definecolor{darkgrey}{rgb}{0.4,0.4,0.4}
      \definecolor{lgray}{rgb}{0.95,0.95,0.95}
      \definecolor{lgreen}{rgb}{0.95,1.00,0.90}
      \definecolor{lred}{rgb}{1.00,0.90,0.80}
      \definecolor{lblue}{rgb}{0.2,0.35,1.00}
      \definecolor{shadecolor}{rgb}{1.00,0.92,0.82}

\def\longlongrightarrow{
\relbar\joinrel\relbar\joinrel\relbar\joinrel\relbar\joinrel\rightarrow}
\def\longlonglongrightarrow{
\relbar\joinrel\relbar\joinrel\relbar\joinrel\relbar\joinrel\relbar\joinrel\relbar\joinrel\rightarrow}

\graphicspath{
	{./figures/}
	{../figures/}
}

\begin{document}

\title{On mass generation in Landau-gauge Yang-Mills theory}

\author{Gernot Eichmann}
\affiliation{LIP Lisboa, Av.~Prof.~Gama~Pinto 2, 1649-003 Lisboa, Portugal}
\affiliation{Departamento de F\'isica, Instituto Superior T\'ecnico, 1049-001 Lisboa, Portugal}
\author{Jan M.~Pawlowski}
\affiliation{ExtreMe Matter Institute EMMI, GSI, Planckstr. 1, 64291 Darmstadt, Germany}
\affiliation{Institut f\"ur Theoretische Physik, Universit\"at Heidelberg, Philosophenweg 16, 69120 Heidelberg, Germany}

\author{Jo\~ao M. Silva}
\affiliation{LIP Lisboa, Av.~Prof.~Gama~Pinto 2, 1649-003 Lisboa, Portugal}
\affiliation{Departamento de F\'isica, Instituto Superior T\'ecnico, 1049-001 Lisboa, Portugal}

\begin{abstract}
      A longstanding question in QCD is the origin of the mass gap in the Yang-Mills sector of QCD, i.e., QCD without quarks. In Landau gauge QCD this mass gap, and hence confinement, is encoded in a mass gap of the gluon propagator, which is found both in lattice simulations and with functional approaches. While functional methods are
       well suited to unravel the mechanism behind the generation of the mass gap, a fully satisfactory answer has not yet been found. In this work we solve the coupled Dyson-Schwinger equations for the ghost propagator, gluon propagator and three-gluon vertex. We corroborate the findings of earlier works, namely that the mass gap generation is tied to the longitudinal projection of the gluon self-energy, which acts as an effective mass term in the equations.
      Because an explicit mass term is in conflict with gauge invariance, this leaves two possible scenarios: If it is
      viewed as an artifact, only the scaling solution survives; if it is dynamical, gauge invariance can only be preserved
      if there are longitudinal massless poles in either of the vertices.
      We find that there is indeed a massless pole in the ghost-gluon vertex, however in our approximation with the assumption of complete infrared dominance of the ghost this pole is only present for the scaling solution.
      We also put forward a possible mechanism that may reconcile the scaling solution, with an infrared dominance of the ghost, with the decoupling solutions based on longitudinal poles in the three-gluon vertex as seen in the PT-BFM scheme.
\end{abstract}

\maketitle

\section{Introduction}

One of the central open questions in strong interaction studies is the origin of mass generation in Quantum Chromodynamics (QCD). One aspect  of the problem is the fact that the majority of the masses of light hadrons, and therefore the visible mass in the universe, must be generated in QCD because light current quarks only carry a small fraction of the mass of the proton. Mass generation in the quark sector is relatively well understood by now in terms of dynamical chiral symmetry breaking and the corresponding dynamical generation of a large quark mass at low momenta, which is seen in various nonperturbative approaches including lattice QCD~\cite{Skullerud:2003qu, Bowman:2005vx, Kizilersu:2006et, Oliveira:2018lln} and functional methods such as Dyson-Schwinger equations (DSEs)~\cite{Cloet:2013jya, Eichmann:2016yit} and the functional renormalization group (fRG)~\cite{Cyrol:2017ewj, Dupuis:2020fhh}.

Another and perhaps more fundamental aspect is the emergence of a mass gap in pure Yang-Mills theory, i.e., QCD without quarks. This is tied to the open question of confinement, and the corresponding problem of mass generation in the Yang-Mills sector of QCD is much less understood.

In principle, the origin of mass generation is encoded in QCD's elementary $n$-point correlation functions. For Yang-Mills theory, these are the two-point functions (the gluon and ghost propagators), three-point functions (the three-gluon vertex and ghost-gluon vertex), four-point functions (e.g., the four-gluon vertex)
and higher $n$-point functions. In particular, it is well-established by now that the massless pole in the perturbative gluon propagator cannot survive in nonperturbative calculations in general covariant gauges including Landau gauge. Moreover, the respective mass gap is directly related to confinement as shown in \cite{Braun:2007bx, Fister:2013bh}.

Possible mechanisms for the generation of the mass gap include the Kugo-Ojima confinement scenario~\cite{Kugo:1979gm},
where the $n$-point functions scale with infrared (IR) power laws~\cite{vonSmekal:1997ohs,vonSmekal:1997ern, Lerche:2002ep, Fischer:2002eq, Fischer:2002hna, Alkofer:2000wg, Zwanziger:2002ia, Pawlowski:2003hq,Alkofer:2008tt} as given in Eq.~\eqref{GZ-scaling} below, a Schwinger mechanism for longitudinal correlation functions~\cite{Aguilar:2006gr, Binosi:2009qm, Rodriguez-Quintero:2010qad,Aguilar:2011xe, Aguilar:2015bud, Aguilar:2016ock, Aguilar:2017dco,Aguilar:2021okw}, and the related gluon condensation mechanism~\cite{Wetterich:1999vd,Gies:2006nz}. In all these scenarios, irregularities in longitudinal and/or transverse projections of the Yang-Mills vertices are triggered and required for the mass gap to be present. While the IR scaling of the correlation functions in the Kugo-Ojima scenario directly induces such irregularities by assuming the existence of BRST charges, these must be triggered explicitly in the Schwinger mechanism and gluon condensation scenarios.

The gluon propagator is parametrized as
  \begin{equation}\label{gluon}
     D^{\mu\nu}(Q) = \frac{1}{Q^2}\left( Z(Q^2)\,T^{\mu\nu}_Q + \xi\,L(Q^2)\,L^{\mu\nu}_Q\right),
   \end{equation}
where
\begin{align}\label{eq:Projections}
T^{\mu\nu}_Q = \delta^{\mu\nu} - \frac{Q^\mu Q^\nu}{Q^2}\,,\qquad  L^{\mu\nu}_Q = \frac{Q^\mu Q^\nu}{Q^2}
\end{align}
 are the transverse and longitudinal projection operators with respect to the four-momentum $Q^\mu$, $\xi$ is the gauge parameter and $\xi=0$ corresponds to Landau gauge. In linear covariant gauges, the longitudinal dressing function $L(Q^2)=1$ is trivial due to gauge invariance, but for later purposes we keep it general in what follows.
A basic mass parameter $m_0$
is defined by the value of the transverse part $D(Q^2) = Z(Q^2)/Q^2$ at $Q^2 \to 0$,
\begin{equation}\label{gluon-prop-0}
     D(Q^2\to 0) \propto \frac{1}{m_0^2}\,.
\end{equation}

\begin{figure}[t]
  \includegraphics[width=0.68\columnwidth]{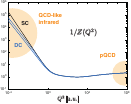}
  \caption{QCD-like behavior of the inverse gluon dressing function $1/Z(Q^2)$ for exemplary scaling (SC)
           and decoupling (DC) solutions.}
  \label{fig-1/z}
  \end{figure}

The IR solutions seen in lattice QCD simulations do not show scaling as required in the Kugo-Ojima confinement scenario and have been called decoupling or massive solutions. In this case, the gluon propagator saturates at low momenta and becomes constant in the IR~\cite{Cucchieri:2008qm, Bogolubsky:2009dc, Maas:2011se, Duarte:2016iko, Aguilar:2019uob} like in Eq.~\eqref{gluon-prop-0}. This solution is also obtained with DSE and  fRG calculations, see e.g.~\cite{Aguilar:2008xm, Boucaud:2008ji,Boucaud:2008ky, Fischer:2008uz,Rodriguez-Quintero:2010qad, Weber:2011nw,Boucaud:2011ug,Aguilar:2015bud, Cyrol:2016tym, Reinosa:2017qtf}.  In addition, functional studies of correlation functions within approximations have revealed the existence of a family of these decoupling solutions, where the maximal decoupling solution is close to that found on the lattice. There are also indications for different decoupling solutions on the lattice depending on the IR details of the gauge-fixing procedure, which involves the removal of Gribov copies~\cite{Sternbeck:2012mf}. In the continuum limit this is a numerically very challenging problem which has not been overcome yet. It has been speculated that the emergence of a family of solutions may be due to an additional gauge fixing parameter in Landau gauge, see e.g.~\cite{Maas:2017csm, Maas:2019ggf}.

Correspondingly, for the decoupling solutions the transverse gluon dressing function $Z(Q^2)$ vanishes like $Z(Q^2)\propto Q^2$ and $1/Z(Q^2) \propto m_0^2/Q^2$ has a singularity at the origin in $Q^2$. This is shown in Fig.~\ref{fig-1/z} and clearly differs from a QED-like behavior where the photon \textit{dressing} function and not the propagator saturates at IR momenta. The origin, details and consequences of this feature are however not yet fully understood. A possible explanation that has been studied in the PT-BFM (Pinch Technique/Background-Field Method) framework is due to the Schwinger mechanism, which could generate longitudinally coupled massless poles in Yang-Mills vertices such as the three-gluon vertex and thereby induce such a behavior in the transverse part of the gluon propagator~\cite{Aguilar:2006gr, Binosi:2009qm, Aguilar:2011xe, Aguilar:2015bud, Aguilar:2016ock, Aguilar:2017dco,Aguilar:2021okw}.

The other endpoint of the family of decoupling solutions for $m_0\to\infty$ is the scaling solution~\cite{vonSmekal:1997ohs,vonSmekal:1997ern, Lerche:2002ep, Fischer:2002eq, Fischer:2002hna, Alkofer:2000wg, Zwanziger:2002ia, Pawlowski:2003hq,Alkofer:2008tt} with the IR scaling
 \begin{equation}\label{GZ-scaling}
      Z(Q^2\to 0) \propto (Q^2)^{2\kappa}\,, \quad
      G(Q^2\to 0) \propto (Q^2)^{-\kappa}\,.
\end{equation}
Here, $G(Q^2)$ is the ghost dressing function and the IR exponent $\kappa$ is typically of the order $\kappa \sim 0.6$ depending on the truncation of the system. Such a behavior is consistent with the Kugo-Ojima confinement scenario based on the assumption of \textit{global} BRST symmetry (existence of BRST charges); see~\cite{Fischer:2008uz,Mader:2013eqo} for detailed discussions.
By contrast, the decoupling scenario entails
\begin{equation}
            Z(Q^2 \to 0) \propto Q^2\,, \qquad
            G(Q^2 \to 0) = const.
\end{equation}

In any case, the gluon propagator is neither gauge invariant nor renormalization-group invariant and thus its value~\eqref{gluon-prop-0}
at vanishing momentum hardly defines a mass gap without further specifications. Instead, a gluon mass gap is best defined as the (spatial) transverse correlation length of this correlation function through the screening mass, see e.g.~\cite{Cyrol:2017qkl} for the finite temperature version. It can be extracted from the gluon propagator $D^{\mu\nu}(Q) $ by a Fourier transform
\begin{align}
 	\label{eq:massgap}
 \lim_{r\to\infty}	\int\frac{d^3 Q}{(2 \pi)^3} \frac{Z(\vect Q^2)}{\vect Q^2}\, e^{i \,\footnotesize \vect x \cdot \vect Q} \propto e^{-  m_\textrm{gap}\,r}
\end{align}
with $r=|\vect x|$. It is the mass parameter $m_\textrm{gap}$ that carries the information about the (physical) mass gap of QCD. For example, it leads to the confinement-deconfinement temperature $T_c(m^E_\textrm{gap},m^M_\textrm{gap})$, where $m^{E,M}_\textrm{gap}$ are the chromo-electric (inverse temporal screening length) and chromo-magnetic (inverse spatial screening length) mass gaps at finite temperature. 
In turn, in particular $m^E_\textrm{gap}$ is also sensitive to the confinement-deconfinement phase transition, see e.g.~\cite{Maas:2011ez,Silva:2016onh}. 
In the vanishing temperature limit both masses reduce to $m_\textrm{gap}$, which is thus  directly linked to confinement, for more details see~\cite{Braun:2007bx,  Fister:2013bh, Cyrol:2017qkl}.

\begin{figure}[t!]
  \includegraphics[width=0.7\columnwidth]{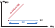}
  \caption{Sketch of the relation between $m_\textrm{gap}$ and $m_0$, with $m_\textrm{gap} \approx$ const.  for the Yang-Mills--like solutions and $m_0 \approx m_\textrm{gap}$ for those in massive Yang-Mills theory. }
  \label{fig-mgap}
  \end{figure}

We emphasize that $m_0$ may change significantly with only minor or no changes inflicted on $m_\textrm{gap}$ for $m_0\gtrsim  m_\textrm{gap}$, and thus the value of $m_0$ does not influence observables such as $T_c$. This is sketched in Fig.~\ref{fig-mgap} (blue branch) and suggests that $m_0$ with $m_0> m_\textrm{gap}$ labels different IR solutions of Yang-Mills theory in Landau-type gauges.  In turn, the red branch in Fig.~\ref{fig-mgap} with $m_0\approx m_\textrm{gap}\to \infty$ clearly differentiates between physically different theories labelled by $m_\textrm{gap}$. Evidently, only solutions to functional equations within the former regime may describe Yang-Mills theory, while the latter regime may be interpreted as solutions in massive Yang-Mills theory with physically different masses. Interestingly, the solution seen on the lattice is close to (or at) the boundary between these two regimes; see~\cite{Cyrol:2016tym} for more details. Note also that the mass gap in the Yang-Mills branch is the minimal one that can be obtained in the system.

We also note that the scaling for the propagators in~\eqref{GZ-scaling} can be embedded in an IR hierarchy for all $n$-point functions, which admits a $ 1/Q^4$ behavior for all diagrams that contribute to the quark-antiquark ($q\bar{q}$) four-point function~\cite{Alkofer:2006gz, Alkofer:2008et}. This leads to a linear rise in the $q\bar{q}$ potential with the distance $r$ in coordinate space already in a single gluon exchange picture. If such a behavior could be shown to be present for  $r \lesssim 1/ m_\textrm{gap}$, it would facilitate the access to observables showing direct signatures of confinement. For the decoupling solutions, on the other hand, no order of a diagrammatic expansion of the $q\bar{q}$ four-point function shows a $1/Q^4$ behavior. However, for the whole family of solutions with $m_0\gtrsim m_\textrm{gap}$, confinement appears at the confinement-deconfinement phase transition with an $m_0$-independent $T_c$~\cite{Braun:2007bx, Fister:2013bh}. Moreover, the expectation value of the Polyakov loop, which is the respective order parameter, can be obtained within a resummation of diagrams~\cite{Herbst:2015ona}.

The open questions we want to address in this work are therefore: What is the mechanism that generates the IR singularities in Fig.~\ref{fig-1/z}? What distinguishes the different decoupling solutions, and is there a preference for one or the other type of solutions?

To study the problem, we solve the coupled DSEs of Landau-gauge Yang-Mills theory for the ghost and gluon propagator and the three-gluon vertex. This builds upon a long history of investigations starting with the DSEs for the two-point functions~\cite{vonSmekal:1997ohs, vonSmekal:1997ern,Alkofer:2000wg, Lerche:2002ep, Fischer:2002eq, Fischer:2002hna} and subsequent improvements regarding the role of three-point functions~\cite{Alkofer:2008dt, Binosi:2011wi, Huber:2012kd, Pelaez:2013cpa, Aguilar:2013vaa, Blum:2014gna, Eichmann:2014xya, Williams:2015cvx, Huber:2017txg, Huber:2018ned, Aguilar:2019uob, Huber:2020keu}, four-point functions~\cite{Kellermann:2008iw, Cyrol:2014kca, Binosi:2014kka, Gracey:2014ola, Eichmann:2015nra, Huber:2017txg, Huber:2018ned, Huber:2020keu}, two-loop contributions~\cite{Meyers:2014iwa, Huber:2017txg, Gracey:2019xom, Huber:2020keu}, the determination of propagators in the complex momentum plane~\cite{Strauss:2012dg, Fischer:2020xnb}, and applications to glueballs~\cite{Sanchis-Alepuz:2015hma, Huber:2020ngt}. Simultaneous advances have been made with fRG calculations~\cite{Ellwanger:1995qf, Ellwanger:1996wy, Bergerhoff:1997cv, Pawlowski:2003hq, Fischer:2004uk, Mitter:2014wpa, Cyrol:2016tym, Cyrol:2017ewj, Dupuis:2020fhh} and in lattice QCD~\cite{Cucchieri:2008qm, Bogolubsky:2009dc, Maas:2011se, Cucchieri:2011ig, Sternbeck:2012mf, Cucchieri:2013nja, Duarte:2016iko, Athenodorou:2016oyh, Maas:2017csm, Aguilar:2019uob}.

In the following we show that the emergence of the IR singularity in Fig.~\ref{fig-1/z} is tied to the longitudinal projection of the gluon self-energy, which we call $\PiT(Q^2)$ and which for the decoupling solutions acts as an effective mass term (see Eq.~\eqref{vac-pol} below). In the present work we offer two possible interpretations: In Scenario A, we interpret this term as an artifact of the regularization and/or truncation; in this case we find that only the scaling solution survives, however with an ambiguity in the IR exponent~$\kappa$. In Scenario B, we consider the term to be dynamical; in this case gauge invariance can only be preserved if there is a massless longitudinal pole either in the ghost-gluon, three-gluon or four-gluon vertex, which drops out from the transverse equations and only serves to eliminate $\PiT(Q^2)$. Here we use the assumption of complete IR dominance of the ghost, and a peculiar finding is the fact that such a pole \textit{does} appear in the ghost-gluon vertex, but \textit{only} for the scaling solution.

The paper is organized as follows. In Sec.~\ref{sec:dses} we discuss the DSEs for the Yang-Mills system in Landau gauge and the different truncations that we employ. In Sec.~\ref{sec:scenario-a} we explain Scenario A and its consequences. In Sec.~\ref{sec:scenario-b} we investigate Scenario B and the emergence of massless longitudinal poles in the ghost-gluon vertex. We conclude in Sec.~\ref{sec:summary}. To keep the paper self-contained, the explicit diagrams appearing in the DSEs are worked out in detail in Appendix~\ref{app:expl}. Appendices~\ref{app:renormalization}, \ref{sec:numerics} and~\ref{sec:long-sing} provide further details on the renormalization, the numerical procedure and the longitudinal poles in the ghost-gluon vertex.
We work in Euclidean conventions throughout the paper.

\begin{figure}[t]
  \includegraphics[width=0.90\columnwidth]{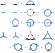}
  \caption{Dyson-Schwinger equations for the ghost propagator (top), gluon propagator (middle) and three-gluon vertex (bottom).
           The gluon self-energy depends on the ghost loop, gluon loop, tadpole, squint and sunset diagrams.
           The three-gluon vertex contains the ghost triangle, gluon triangle and swordfish diagrams,
           plus further two-loop terms and diagrams with higher $n$-point functions (not shown).
           The ghost-gluon vertex (green) and four-gluon vertex (orange) satisfy their own equations and are external inputs in the system. }
  \label{fig-ym}
\end{figure}

\section{Yang-Mills DSEs}\label{sec:dses}

In the present work we aim at a thorough understanding of the mechanisms at work within the mass generation in Yang-Mills theory in Landau gauge QCD in a functional formulation. While quantitatively reliable approximations have been set up in functional approaches, see e.g.~\cite{Cyrol:2016tym, Huber:2020keu}, we shall employ  approximations here that carry all the dynamics of the system but are still simple enough to access these dynamics directly.

Accordingly, we solve the coupled DSEs for its lowest $n$-point functions shown in Fig.~\ref{fig-ym}: the ghost propagator, gluon propagator and the three-gluon vertex. The propagator equations are exact since the gluon DSE also includes the two-loop terms. In the three-gluon vertex DSE we neglect further two-loop terms and diagrams including vertices without a tree-level counterpart. To keep the discussion transparent, we relegate the explicit formulas to Appendix~\ref{app:expl} and only highlight the most important aspects in the main text.

Throughout this work we keep the ghost-gluon vertex at tree level because this is a good approximation in Landau gauge~\cite{Taylor:1971ff,vonSmekal:1997ern}: Here the vertex is finite in the ultraviolet (UV) and does not need to be renormalized, and explicit calculations have shown that the deviations from the tree-level behavior are small~\cite{Huber:2012kd,Huber:2020keu}. The three-gluon vertex, on the other hand, is almost completely dominated by its classical tensor structure and has only a mild angular dependence~\cite{Eichmann:2014xya}. Thus, in the following we restrict ourselves to the leading tensor and the symmetric limit  such that the vertex is represented by a one-dimensional function $F_{3g}(Q^2)$. The quantities we compute are thus the ghost dressing $G(Q^2)$, the gluon dressing $Z(Q^2)$ and the three-gluon vertex dressing $F_{3g}(Q^2)$.

We consider the following truncations of the coupled DSEs in Fig.~\ref{fig-ym}:\\[-2ex]

{\tiny$\blacksquare$}
\textbf{Setup 1:}
This is the DSE system solved in~\cite{Lerche:2002ep,Fischer:2002eq,Fischer:2002hna}, where the three-gluon vertex DSE is bypassed and only the propagator equations are solved,  also neglecting the two-loop terms in the gluon DSE.    To avoid introducing model input for the three-gluon vertex, one can  use the ansatz $F_{3g}(Q^2)=G(Q^2)/Z(Q^2)$, which ensures the correct renormalization and perturbative limit of the vertex, or keep the vertex at tree level with $F_{3g}(Q^2)=const.$  For our purposes, Setup 1 will mainly serve as a reference point since  it is well established by now that the three-gluon vertex is suppressed at small momenta and  likely has a zero crossing~\cite{Cucchieri:2008qm, Maas:2011se, Huber:2012kd, Pelaez:2013cpa, Aguilar:2013vaa, Eichmann:2014xya, Athenodorou:2016oyh, Huber:2018ned, Aguilar:2019uob}.\\[-2ex]

{\tiny$\blacksquare$}
\textbf{Setup 2:} Here we still neglect the two-loop terms in the gluon DSE  but back-couple the three-gluon vertex DSE into the system.  For the four-gluon vertex, which now appears as an additional input,  we employ its classical tensor multiplied with $G(Q^2)^2/Z(Q^2)$, which again  ensures the correct renormalization of the vertex and its perturbative limit.  We will see below that the error induced by this truncation is at the 10\% level.\\[-2ex]

{\tiny$\blacksquare$}
\textbf{Setup 3:} This corresponds to the full system in Fig.~\ref{fig-ym}  also including the two-loop terms in the gluon DSE.    As a consequence, the propagator DSEs are two-loop complete at UV momenta.  The remaining inputs are the ghost-gluon and four-gluon vertex  where we use the ansätze discussed above.  Below we will see that the error in this truncation is at the 3--4\% level.\\[-2ex]

We emphasize that there is no explicit model input in any of these truncations (except for dropping higher vertices and tensor structures) and  the only parameter in all cases is the coupling $g$.   Furthermore, the qualitative features found in this work  are independent of the truncations and appear in all three setups.

The general form of the gluon DSE is given by
\begin{subequations}\label{eq:DSE0}
\begin{equation}\label{dses-0}
(D^{-1})^{\mu\nu}(Q) = (D_0^{-1})^{\mu\nu}(Q) + \Pi^{\mu\nu}(Q)\,,
\end{equation}
where
\begin{equation}\label{vac-polTrans}
	\Pi^{\mu\nu}(Q) = \mathbf\Pi_\textrm{phys}(Q^2)\,Q^2\,T^{\mu\nu}_Q\,,
\end{equation}
\end{subequations}
with the transverse projection operator defined in \eqref{eq:Projections}. In \eqref{dses-0}, the tree-level propagator $D^{\mu\nu}_0(Q)$ is given by Eq.~\eqref{gluon} with the replacement $Z(Q^2) \to 1/Z_A$, where $Z_A$ is the gluon renormalization constant. The gluon self-energy $\Pi^{\mu\nu}(Q)$ in \eqref{dses-0} is the sum of the ghost loop, gluon loop, tadpole, squint and sunset diagrams in Fig.~\ref{fig-ym}. The transversality of the gluon self energy in \eqref{gluon} and \eqref{eq:DSE0} is a direct consequence of the  Slavnov-Taylor identities (STIs) for general covariant gauges, e.g.~\cite{Napetschnig:2021ria}.

In approximations transversality might be lost. Indeed, already in perturbation theory the transversality of \eqref{vac-polTrans} is not present in the single diagrams in Fig.~\ref{fig-ym}, and longitudinal pieces that are related by gauge symmetry cancel between diagrams. This is used in the PT-BFM scheme, together with dimensional regularisation, for a reorganization of classes of diagrams according to transversality; see e.g.~\cite{Aguilar:2015bud}.

Hence, in order to account for truncation artifacts we allow for a more general form of the gluon self energy,
\begin{align}\label{eq:GenericPi}
	\Pi^{\mu\nu}(Q) = \Delta_T\,Q^2\,T^{\mu\nu}_Q +\Delta_0 \,\delta^{\mu\nu} +\Delta_L\,Q^2\, L^{\mu\nu}_Q\,,
\end{align}
as already done in \eqref{gluon}, where we allowed for a momentum-dependent longitudinal dressing $L(Q)$ for the propagator. Note that  $\Delta_0(Q^2)$ and $Q^2 \Delta_L(Q^2) $ decay for large momenta $Q^2\gg \Lambda^2$, see e.g.~\cite{Cyrol:2016tym, Dupuis:2020fhh, mSTI2021}.

Inserting Eqs.~\eqref{gluon} and~\eqref{eq:GenericPi} in~\eqref{dses-0} and comparing coefficients leads to the following equations for the propagator dressing functions $G(Q^2)$, $Z(Q^2)$ and $L(Q^2)$,
\begin{equation} \label{eq:self-energy-GenericPi}
\begin{split}
	G(Q^2)^{-1} &= Z_c + \Sigma(Q^2)\,, \\[1ex]
	Z(Q^2)^{-1} &= Z_A + \Delta_T(Q^2)+ \frac{\Delta_0(Q^2)}{Q^2}\,,  \\[1ex]
	L(Q^2)^{-1} &= 1 + \xi\, \left( \Delta_L(Q^2)+\frac{\Delta_0(Q^2)}{Q^2} \right)\,.
	\end{split}
\end{equation}
These relations hold in all linear covariant gauges.
The set of DSEs in \eqref{eq:self-energy-GenericPi} makes it apparent that the tensor basis used in \eqref{eq:GenericPi} is over-complete: one of the $\Delta$'s can be reabsorbed in a redefinition of the other two. In the absence of approximations, \eqref{vac-polTrans} holds and we arrive at
\begin{align} \label{eq:Trans+LongConsistency}
\mathbf\Pi_\textrm{phys} = \Delta_T+\frac{\Delta_0}{Q^2}\,,\qquad  \Delta_L = - \frac{\Delta_0}{Q^2}\,.
\end{align}
We introduced the over-complete basis~\eqref{eq:self-energy-GenericPi} because there are two qualitatively different sources for the $\Delta$'s, which is important for the discussion of the IR behavior:\\[-2ex]

{\tiny$\blacksquare$}  In numerical applications of functional approaches, loop integrals are regularized by a momentum cutoff $\Lambda$, where loop momenta $k^2>\Lambda^2$ are dropped. The respective renormalization as well as the related Bogolyubov-Parasiuk-Hepp-Zimmermann (BPHZ) renormalization requires mass counter terms for guaranteeing the cutoff independence and transversality of~\eqref{vac-polTrans}, see also~\cite{Gao:2021wun}. Potential remnants are proportional to $\delta^{\mu\nu}$ and may lead to $\Delta_0$. Accordingly, a consistent treatment of these artifacts
is of eminent importance for avoiding explicit mass terms in the IR. \\[-2ex]

{\tiny$\blacksquare$}  Any truncation in which tensor structures of vertices are dropped may lead to artifacts,
which can be distributed between $\Delta_T$ and $\Delta_L$ due to the over-completeness of the basis.
While the $\Delta_T$ contribution is simply a correction to $\mathbf\Pi_\textrm{phys}$, the generation of $\Delta_L$ is potentially harmful in the IR.

The generation of $\Delta_L$ can be elucidated at the relevant example of the ghost-gluon vertex $\Gamma^\mu_\textrm{gh}$. Its complete form has two tensor structures,
\begin{equation}\label{ggl-vertex-0}
	\Gamma^\mu_\text{gh}(p,Q) = -ig  f_{abc}\left[ (1+A)\,p^\mu + B\,Q^\mu\right],
\end{equation}
where $p^\mu$ is the outgoing ghost momentum and $Q^\mu$ the incoming gluon momentum.
In Landau gauge, the ghost renormalization contant $\widetilde Z_\Gamma = 1$
so that the dressing functions $A(p^2,p\cdot Q,Q^2)$ and $B(p^2,p\cdot Q, Q^2)$
measure the deviation from the classical vertex.
In our present approximation we set $A=0$. Evidently, if we also set $B=0$ we would remove a completely longitudinal part of the ghost diagram in the gluon self energy in Fig.~\ref{fig-ym} adding to $\Delta_L(Q^2)$. This entails that the cancellation of all longitudinal parts in the sum of diagrams will be absent and the self-energy will have longitudinal parts.  Moreover, as $B$ drops for large momenta, so will the longitudinal part inflicted by this approximation. Accordingly, this specific approximation artifact does not affect multiplicative renormalization and in particular does not contribute to the mass counter term.\\[-2ex]

In summary, the regularization with a momentum cutoff as well as truncations may lead to artifacts in the gluon DSE that may complicate the identification of the transverse part, in particular in the IR.

In numerical applications of functional approaches, the over-complete basis in \eqref{eq:GenericPi} is usually not used; instead one  absorbs either $\Delta_0$ or $\Delta_L$ in  two other dressing functions. We denote these by
\begin{equation}\label{self-energy-delta-2}
	\begin{split}
		\mathbf\Pi(Q^2) &= \Delta_T(Q^2) + \frac{\Delta_0(Q^2)}{Q^2}\,, \\
		\Pi(Q^2) &= \Delta_T(Q^2) - \Delta_L(Q^2)\,, \\[1mm]
		\frac{\PiT(Q^2)}{Q^2} &= \Delta_L(Q^2) + \frac{\Delta_0(Q^2)}{Q^2}
	\end{split}
\end{equation}
with $\mathbf\Pi(Q^2)=\Pi(Q^2)+\widetilde{\Pi}(Q^2)/Q^2$,
which lead to the following self-energy decompositions:\\[-2ex]

{\tiny$\blacksquare$}
$\Pi(Q^2)$ and $\PiT(Q^2)$ correspond to the decomposition
\begin{equation}\label{vac-pol}
	\Pi^{\mu\nu}(Q) = \Pi(Q^2)\,Q^2\,T^{\mu\nu}_Q  + \widetilde{\Pi}(Q^2)\,\delta^{\mu\nu}\,,
\end{equation}
which we employ for the explicit formulas in App.~\ref{app:expl}.
It facilitates the consistent treatment of the UV renormalization by keeping the $\delta^{\mu\nu}$ term explicit,
and it is convenient because $\Pi(Q^2)$ and $\PiT(Q^2)$ are free of kinematic $1/Q^2$ singularities.
For the decoupling solutions $\Delta_0(0)$ becomes constant and $\Delta_T(Q^2 \to 0)$ has a logarithmic divergence
stemming from the ghost loop.
Note, however, that within approximations the transverse part $\Pi = \Delta_T - \Delta_L$ also contains the purely longitudinal contribution $\Delta_L$, which can come from missing tensor structures in vertices such as the longitudinal part $B$ of the ghost-gluon vertex in \eqref{ggl-vertex-0}. A consistent treatment of these terms  may be of importance for the generation of the mass gap.\\[-2ex]

{\tiny$\blacksquare$} The transverse-longitudinal decomposition
\begin{equation}\label{vac-pol-TL}
	\Pi^{\mu\nu}(Q) = \mathbf\Pi(Q^2)\,Q^2\,T^{\mu\nu}_Q + \widetilde{\Pi}(Q^2)\,L_Q^{\mu\nu} \,,
\end{equation}
on the other hand, corresponds to the DSEs for $Z(Q^2)$ and $L(Q^2)$ in~\eqref{eq:self-energy-GenericPi}.
In this case, the transverse part $\mathbf\Pi(Q^2)$ includes potential artifacts of the UV renormalization,
which must be treated consistently to avoid the implicit introduction of an explicit gluon mass.\\[-2ex]
\begin{figure}[t]
  \includegraphics[width=1\columnwidth]{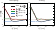}
  \caption{Contributions to the total gluon self-energy $\mathbf{\Pi}(Q^2)$ for decoupling and scaling solutions
           with $\lambda=1$, cf.~Eq.~\eqref{projectors}.
           For each contribution we plot asinh\,$\Pi(Q^2)$ since these can change signs and diverge in the infrared.}
  \label{fig-Zinv}
  \end{figure}

As an example, we consider a numerical solution of the DSEs projected on the transverse part by contracting \eqref{vac-pol} or \eqref{vac-pol-TL} with a transverse
projection operator, hence computing the dressing $\mathbf\Pi(Q^2)=\Delta_T(Q^2) + \Delta_0(Q^2)/Q^2$.
In Fig.~\ref{fig-Zinv} we show exemplary decoupling and scaling solutions obtained in the present work for $\mathbf{\Pi}(Q^2)$  and their breakdowns into individual diagrams. The ghost loop contribution is positive, the gluon loop negative and the two-loop terms only have a small effect.

Interestingly, $\Delta_0/Q^2$ drops rapidly and already vanishes for values of $Q^2$ in the mid-momentum region,
which entails that the UV renormalization is done consistently. In turn, $\Delta_0/Q^2$ contributes to the IR behavior. In fact, in the decoupling case the $1/Q^2$ singularity in the inverse gluon dressing and hence the gluon mass gap is solely produced by this term, which is tantamount to an explicit mass. In the scaling case, the term is still present but has a steeper divergence induced by the ghost loop which becomes large. As a consequence, both $\Delta_0/Q^2$ and $\Delta_T$ diverge with the same power in the IR,  which leads to the IR ghost dominance with the scaling behavior $1/Z(Q^2\to 0) \propto (Q^2)^{-2\kappa}$.

Clearly, all these points need to be understood and the consistent definition of the transverse dressing is of eminent importance for the correct description of the generation of the gluon mass gap. Hence, in the following we shall consider two scenarios: \\[-2ex]

{\tiny$\blacksquare$}
	\textbf{Scenario A:} We assume $\Delta_L \equiv 0$, i.e., we drop terms that are purely longitudinal. Therefore, we must  dispose of $\Delta_0$ as well
to ensure $\PiT = 0$. With this assumption we can extract the respective transverse dressing by an appropriate projection. This is discussed in detail in Sec.~\ref{sec:scenario-a}. \\[-2ex]

{\tiny$\blacksquare$}  \textbf{Scenario B:} We assume $\Delta_L\neq 0$, in which case we need $\Delta_L = -\Delta_0/Q^2$ for consistency, see \eqref{eq:Trans+LongConsistency}. Then, the transverse dressing is simply obtained by contracting the DSE with $T_Q^{\mu\nu}$. 
This is discussed in detail in Sec.~\ref{sec:scenario-b} and  entails a dynamical generation of the mass gap. \\[-2ex]

We shall see that the generation of the gluon mass gap in Scenario A, in the present approximation, is not dynamical but rather similar to that in massive Yang-Mills theory. In turn, the generation of the gluon mass gap in Scenario B is indeed dynamical.
While this entails a clear preference for Scenario B, the discussion in Sec.~\ref{sec:scenario-a}
will serve to introduce the concepts and provide the technical details, which are the same in both scenarios
along with many of the results.

We also note that there are no ambiguities in the practical extraction of any set of two linearly independent dressing functions.
The only purely longitudinal part proportional to $\Delta_L$ that can appear in our approximations arises from the
longitudinal tensor in the ghost-gluon vertex, which we treat as an `extra' term in Sec.~\ref{sec:scenario-b}.
Thus, for the explicit formulas in Appendix~\ref{app:expl} without this term, which are based on the decomposition~\eqref{vac-pol},
one can identify $\Pi = \Delta_T$ and $\PiT = \Delta_0$.

\section{Gluon mass gap: Scenario A} \label{sec:scenario-a}

Scenario A, as defined at the end of the last section, is based on the assumption that purely longitudinal terms are absent, i.e., $\Delta_L(Q^2) = 0$ in Eq.~\eqref{eq:GenericPi}. Then,  $\widetilde{\Pi}(Q^2) = \Delta_0(Q^2)$ has to be an artifact either stemming from the hard cutoff employed in the DSEs and/or from the truncation of the equations. Thus, in the full system with a gauge-invariant regulator or with appropriately defined counter terms it would vanish identically. In this way, systematically improving the truncations would improve the precision on $\Pi(Q^2)$ while eventually sending $\widetilde\Pi(Q^2)$ to zero, so we might as well drop it right away.

 A way to interpolate between the two cases with and without this term is to contract the self-energy with the general projection operator
\begin{equation}\label{projector-general}
 	P^{\mu\nu}_Q = \frac{T_Q^{\mu\nu} - 3 (1-\lambda)\,L^{\mu\nu}_Q}{3Q^2}
\end{equation}
with a parameter $\lambda$, where the transverse and longitudinal projection operators $T_Q^{\mu\nu}$ and $L^{\mu\nu}_Q$ have been defined in \eqref{eq:Projections}. Contracting \eqref{vac-pol} or~\eqref{vac-pol-TL} with \eqref{projector-general} leads to
\begin{equation}\label{projectors}
	\begin{split}
		P^{\mu\nu}_Q\,\Pi^{\mu\nu}(Q) &= \Pi(Q^2) + \lambda\,\frac{\widetilde{\Pi}(Q^2)}{Q^2}\,, \\
		L^{\mu\nu}_Q\,\Pi^{\mu\nu}(Q) &= \widetilde{\Pi}(Q^2)
	\end{split}
\end{equation}
where $\lambda=1$ corresponds to the transverse projection and $\lambda=0$ is the Brown-Pennington projection~\cite{Brown:1988bn}.
\begin{figure}[t]
	\includegraphics[width=0.55\columnwidth]{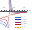}
	\caption{Determination of the infrared exponent $\kappa$ for a tree-level ghost-gluon vertex from the condition~\eqref{kappa-determination}.}
	\label{fig-ir-exponent}
\end{figure}

To begin with, for $\lambda=0$ one cannot find QCD-like decoupling solutions, or even convergent DSE solutions. This is already apparent in the numerical solution in Fig.~\ref{fig-Zinv}. There, the mass gap and $m_0$ solely originate from  $\widetilde\Pi$. In its absence, the system would have to exhibit a Coulomb-type solution which is not present.

In turn, the scaling solution in principle still exists, but for $\lambda=0$ the analytic IR analysis leads to a 0/0 problem~\cite{Lerche:2002ep, Fischer:2002eq, Reinosa:2017qtf}, which is shown in Fig.~\ref{fig-ir-exponent}. For a tree-level ghost-gluon vertex, the algebraic condition that determines the IR exponent $\kappa$ reads
\begin{equation}\label{kappa-determination}
	1 + \lambda\,\frac{4\kappa-3}{2\,(1-2\kappa)} - \frac{6\,(3-2\kappa)}{(1+\kappa)(2+\kappa)} \stackrel{!}{=} 0\,.
\end{equation}
For a transverse projection ($\lambda=1$) this yields $\kappa \approx 0.595$.
Sending $\lambda \to 0$ entails $\kappa \to {1}/{2}$, but for $\lambda=0$
the solution disappears because $\lambda/(1-2\kappa) = 0/0$.

The above considerations suggest to regularize the limit $\PiT\to 0$: We replace it by an auxiliary mass term, which is finally sent to zero in a controlled way. Thus, we investigate the situation with a \textit{constant} mass term
\begin{equation}\label{PiT-const}
	\PiT(Q^2) = \frac{g^2}{4\pi}\,G_0^2 \,\beta \mu^2\,.
\end{equation}
Here, $g$ is the coupling and $G_0 = G(Q^2=0)$ is the value of the ghost dressing function at the origin. This ensures the correct renormalization as will become clear below. Furthermore, $\beta$ is a dimensionless mass parameter and $\mu$ is the renormalization scale. In this way, the equations explicitly depend on a mass parameter $\beta$ which is sent to zero in the end. Note that this is formally similar to massive Yang-Mills theory, i.e., the Curci-Ferrari-model which has been studied in a series of works, e.g.~\cite{Reinosa:2017qtf,Gracey:2019xom, Pelaez:2021tpq}. However, in our case the mass parameter $\beta$ does not arise from the Lagrangian; it is merely an artifact which must be taken to zero to recover massless Yang-Mills theory. Note that Eq.~\eqref{PiT-const} is also a conceptual simplification since there is no need to deal with quadratic divergences.

   \subsection{Renormalization}\label{sec:renormalization-main}

What can  obscure the study of DSEs to some extent is their dependence on the renormalization constants or, equivalently, on the renormalized values of the gluon and ghost dressing functions $Z(\mu^2) = Z_\mu$ and $G(0) = G_0$. Hence, before we proceed with the explicit numerical solutions in Section~\ref{ScA-solutions}, we discuss the renormalization of the Yang-Mills system in detail.
We note that the following discussion is independent of the two scenarios and  equally applies to Scenario B.

Propagators, vertices and the coupling are related to their bare counterparts, denoted by the superscript $(B)$, by the multiplicative renormalization constants,
\begin{equation} \label{ren-consts}\renewcommand{\arraystretch}{1.4}
	\begin{array}{rl}
		G^{(B)} &= Z_c\,G\,, \\[1ex]
		Z^{(B)} &= Z_A \,Z\,, \\[1ex]
		g^{(B)} &= Z_g\,g\,,
	\end{array}\qquad
	\begin{array}{rl}
		\Gamma_\text{gh} &= \widetilde Z_\Gamma\,\Gamma_\text{gh}^{(B)}\,, \\[1ex]
		\Gamma_\text{3g} &= Z_{3g}\,\Gamma_\text{3g}^{(B)}\,, \\[1ex]
		\Gamma_\text{4g} &= Z_{4g}\,\Gamma_\text{4g}^{(B)}\,,
	\end{array}
\end{equation}
where $\widetilde Z_\Gamma = Z_g\,Z_A^{1/2}\,Z_c$, $Z_{3g} = Z_g\,Z_A^{3/2}$ and $Z_{4g} = Z_g^2\,Z_A^2$ as a consequence of the STIs. In the Landau gauge the ghost-gluon vertex stays unrenormalized, so we can set $\widetilde Z_\Gamma=1$. Therefore, all renormalization constants can be related to $Z_A$ and $Z_c$,
\begin{equation}\label{ren-constants-2}
	Z_g = \frac{1}{Z_A^{1/2}\,Z_c}\,, \qquad
	Z_{3g} = \frac{Z_A}{Z_c}\,, \qquad
	Z_{4g} = \frac{Z_A}{Z_c^2}\,.
\end{equation}
We can consistently renormalize the ghost and gluon propagators at different renormalization points: In practice it is convenient to renormalize the gluon dressing function at $Q^2=\mu^2$ to $Z(\mu^2) = Z_\mu$, while the ghost dressing is fixed at $Q^2=0$ to $G(0) = G_0$. This fixes all renormalization constants and the resulting equations depend on $g$, $Z_\mu$ and $G_0$, see Appendix~\ref{app:renormalization} for details:
\begin{equation}\label{dses-1}
	\begin{split}
		G(Q^2)^{-1} &= G_0^{-1} + \Sigma(Q^2) - \Sigma(0)\,, \\[1ex]
		Z(Q^2)^{-1} &= Z_\mu^{-1} + \mathbf{\Pi}(Q^2) - \mathbf{\Pi}(\mu^2)\,, \\[1ex]
		F_{3g}(Q^2) &= Z_{3g} + \mM(Q^2)\,.
	\end{split}
\end{equation}
Here, $\Sigma$ is the ghost self-energy, $\mathbf{\Pi} = \Pi + \PiT/Q^2$ the gluon self-energy in Eq.~\eqref{vac-pol-TL},  and $\mM$ are the vertex diagrams. The renormalization constants are dynamically determined by
\begin{equation}\label{rcs-1}
	Z_c = G_0^{-1} - \Sigma(0)\,, \quad
	Z_A = Z_\mu^{-1} - \mathbf{\Pi}(\mu^2)\,,
\end{equation}
which fixes $Z_{3g}$ and $Z_{4g}$ from Eq.~\eqref{ren-constants-2}
so that no further renormalization of the vertices is required.

\begin{figure}[t]
	\includegraphics[width=0.9\columnwidth]{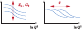}
	\caption{Renormalization and scale setting in the Yang-Mills system: $Z_\mu$ and $G_0$ renormalize the propagators whereas $g$ sets the scale.}
	\label{fig-renormalization}
\end{figure}

In practical solutions of the Yang-Mills DSEs one keeps $g$ and $Z_\mu$ fixed;
the value $G_0$ of the ghost dressing at vanishing momentum then distinguishes the scaling and decoupling solutions.
Any finite $G_0$ corresponds to a decoupling solution and the limit $G_0\to\infty$ to the scaling solution.
This is, however, not completely satisfactory from the viewpoint of renormalization as sketched in Fig.~\ref{fig-renormalization}.
$Z_\mu$ and $G_0$ should only multiplicatively renormalize the propagators, which would not lead to physically different solutions;
in a double-logarithmic plot, a renormalization only induces vertical shifts in both functions.
Similarly, the value of the coupling $g$ should only set the scale, and when plotted over a logarithmic momentum scale
it would only induce horizontal shifts in both functions. But how is it then possible to obtain a \textit{family} of
different decoupling solutions?

To this end, let us redefine the renormalized propagators by dividing out their values at the respective renormalization scales,
\begin{equation}
	G(Q^2) \to \frac{G(Q^2)}{G_0}\,,\qquad
	Z(Q^2) \to \frac{Z(Q^2)}{Z_\mu}\,,
\end{equation}
with the same redefinition for all other quantities that renormalize like $G$ and $Z$, i.e., the renormalization constants
\begin{equation}
	Z_c \to Z_c\,G_0\,, \qquad
	Z_A \to Z_A\,Z_\mu
\end{equation}
and the three- and four gluon vertices
\begin{equation}
	F_{3g} \to F_{3g}\,\frac{Z_\mu}{G_0}\,, \qquad
	F_{4g} \to F_{4g}\,\frac{Z_\mu}{G_0^2}\,.
\end{equation}
This is always possible since the dressing functions are only fixed up to multiplicative renormalization.
For the scaling solution, $G_0$ becomes infinite. However, throughout this work
as well as in most of the literature, the scaling case is defined as the \textit{limit} of the decoupling solutions for $G_0 \to \infty$. Then, all redefinitions are well-defined.

By inspection of the DSEs, see Appendix~\ref{app:renormalization}, one finds
that $g$, $Z_\mu$ and $G_0$ no longer appear individually in the equations but only in the combination
\begin{equation}\label{def-alpha}
	\alpha := \frac{g^2}{4\pi}\,Z_\mu\,G_0^2\,.
\end{equation}
Henceforth, we call $\alpha$ the \textit{coupling} parameter, which can take any positive value.
Note that $\alpha$ should not be confused with the strong coupling $g^2/(4\pi)$; it has the $\mu$ running of the gluon propagator
and will later be related to the mass parameter.
The equations~(\ref{dses-1}--\ref{rcs-1}) with redefined quantities take the form
\begin{equation}\label{dses-redef-1}
	\begin{split}
		G(Q^2)^{-1} &= 1 + \alpha\left[\Sigma(Q^2) - \Sigma(0)\right], \\[1ex]
		Z(Q^2)^{-1} &= 1 + \alpha\left[\mathbf{\Pi}(Q^2) - \mathbf{\Pi}(\mu^2)\right], \\[1ex]
		F_{3g}(Q^2) &= Z_{3g} + \alpha\,\mM(Q^2)\,, \\[1ex]
		Z_c &= 1 - \alpha\,\Sigma(0)\,, \\[1ex]
		Z_A &= 1 - \alpha\,\mathbf{\Pi}(\mu^2)\,.
	\end{split}
\end{equation}
Here we have extracted the factor $\alpha$ from the redefined self-energies,  which do not explicitly depend on it, apart from the two-loop terms. Thus, all renormalization constants have disappeared and only the coupling $\alpha$ remains.

With Eq.~\eqref{def-alpha} it becomes clear that it is technically not $G_0$ that distinguishes the decoupling solutions but $\alpha$: If $g$ and $Z_\mu$ are kept fixed, changing $G_0$ is equivalent to changing $\alpha$, but equivalently we could have kept any other two variables fixed and changed the third variable. Thus, any finite $\alpha$  produces a decoupling solution and $\alpha\to\infty$ corresponds to the scaling solution.

It is also convenient to remove the explicit dependence on the renormalization scale $\mu$.
To do so, we introduce a dimensionless variable $x$,
\begin{equation}\label{rescaling}
	x = \frac{Q^2}{\beta\mu^2}\,,
\end{equation}
which is always possible since the momentum scale still carries arbitrary units and only rescales
the dressing functions according to Fig.~\ref{fig-renormalization}. If we perform the same operation
for the loop momenta inside the integrals, then the dependence on $\mu^2$ drops out from the equations
and  moves to the cutoff of the integrals.
The mass parameter $\beta$ disappears from $\PiT(Q^2)$  as well,
which in the redefined system becomes
\begin{equation}\label{PiT-redef}
	\frac{\PiT(Q^2)}{Q^2} \to \frac{1}{x} \,.
\end{equation}
In turn, $\beta$ appears in the subtraction point since $Q^2=\mu^2$ entails $x=1/\beta$. To give a concrete example, in the redefined system the ghost self-energy from Eq.~\eqref{ghost-self-energy-LG} in Appendix~\ref{app:expl} reads
\begin{equation}\label{ghost-self-energy-redeff}
	\Sigma(x) = -\frac{N_c}{2\pi^2}\int_0^L du\,u^2 \int_{-1}^1  dz\,(1-z^2)^{\frac{3}{2}}\,\frac{G(u_+)\,Z(u_-)}{u_+ u_-^2}\,.
\end{equation}
Here, $u=k^2/(\beta \mu^2)$, $u_\pm = u + x/4 \pm \sqrt{u x}\,z$,
and the $\mu^2$ dependence has moved into the cutoff $L$.

Finally, we redefine the ghost dressing function, and every quantity that renormalizes with it, once more by
\begin{equation}\label{ghost-redef-2}
	G(x) \to \sqrt{\alpha}\,G(x)\,.
\end{equation}
As a consequence, the coupling $\alpha$ no longer appears in front of the self-energies but only in the ghost DSE. Then the renormalized equations
are given by
\begin{equation}\label{dses-redef-2}
	\begin{split}
		G(x)^{-1} &= \frac{1}{\sqrt{\alpha}} + \Sigma(x) - \Sigma(0)\,, \\[1ex]
		Z(x)^{-1} &= 1 + \mathbf{\Pi}(x) - \mathbf{\Pi}(\tfrac{1}{\beta})\,, \\[1ex]
		F_{3g}(x) &= Z_{3g} + \mathcal{M}(x)\,, \\[1ex]
		Z_c &= \frac{1}{\sqrt{\alpha}} - \Sigma(0)\,, \\[1ex]
		Z_A &= 1 - \mathbf{\Pi}(\tfrac{1}{\beta})
	\end{split}
\end{equation}
with $\mathbf{\Pi}(x) = \Pi(x) + 1/x$. This is equivalent to the original equations but now the only explicit parameters are $\alpha$ and $\beta$. The subtraction point is no longer arbitrary but tied to the mass parameter $\beta$. In particular, the massless limit $\beta\to 0$ corresponds to a subtraction at $x\to\infty$. Here it is also obvious why $\alpha\to\infty$ corresponds to the scaling solution, because $\alpha$ only appears in the ghost DSE and nowhere else, with $G(0) = \sqrt{\alpha}$. Note that we could have equally absorbed the coupling in the gluon dressing by setting $Z(x) \to \alpha Z(x)$. This would lead to $G(0)=1$ and $Z(1/\beta) = \alpha$ but for $0 < \alpha < \infty$ the solutions are always identical up to multiplicative factors.

At this point, the scale $x$ in all  equations above is still arbitrary and given in internal units. A quantity that is invariant under renormalization is the running coupling
\begin{equation}\label{running-coupling-0}
	\conjg\alpha(x) = Z(x)\,G(x)^2\,,
\end{equation}
which remains finite even for $\alpha\to\infty$. Here the parameter $\alpha$ is absorbed in the ghost dressing by Eq.~\eqref{ghost-redef-2}. In dynamical QCD, the scale should be fixed to experiment by setting the value of $\conjg\alpha(x)$ at some momentum scale, which sets the scale in GeV units. In Yang-Mills theory, the usual procedure is to set the scale by comparing with quenched lattice QCD.

\begin{figure*}
	\begin{center}
		\includegraphics[width=0.72\textwidth]{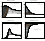}
	\end{center}
	\vspace{-5mm}
	\caption{Solutions for the running coupling and the ghost, gluon and three-gluon vertex dressing functions in Setup 3
		for $\beta=0.001$, $c=0.95$ and $0.003 < \alpha < 10^{16}$.
		We rescaled the curves such that for each $\alpha$ the maximum of the gluon dressing appears at $x=1$.
		The orange bands indicate the onset of the decoupling solutions for $0.6 \lesssim \alpha \lesssim 1$.}\label{fig:results-S2}
\end{figure*}
At asymptotically large momenta, the three- and four-gluon vertex dressings satisfy
\begin{equation}\label{couplings-uv}
	F_{3g}(x) \to \frac{G(x)}{Z(x)}\,, \qquad
	F_{4g}(x) \to \frac{G(x)^2}{Z(x)}\,.
\end{equation}
Hence, we can define equivalent `running couplings' by multiplying $\conjg\alpha(x)$ with powers of the renormalization-group invariants $Z(x) F_{3g}(x)/G(x)$ and $Z(x) F_{4g}(x)/G(x)^2$, e.g.,
\begin{equation}\label{running-couplings-0}
	\begin{split}
		\conjg\alpha_{3g}(x) &= Z(x)^3\,F_{3g}(x)^2\,, \\[1ex]
		\conjg\alpha_{4g}(x) &= Z(x)^2\,F_{4g}(x)\,.
	\end{split}
\end{equation}
At large momenta these couplings are all identical but in the IR they can be different. In our calculations we set the four-gluon vertex to $F_{4g}(x) = G(x)^2/Z(x)$, so the second relation in~\eqref{couplings-uv} is trivially satisfied. The deviation from the first relation will serve as a measure of the truncation error, which we discuss below.

\subsection{Discussion of solutions}\label{ScA-solutions}

We proceed with the numerical solution of the DSEs~\eqref{dses-redef-2} in the $(\alpha,\beta)$ plane, where $\alpha$ is the coupling and $\beta$ the mass parameter as explained above. The self-energy contributions for the ghost, gluon and three-gluon vertex DSEs are worked out in Appendix~\ref{app:expl}.
For the gluon DSE in Scenario A, we only need to keep the terms contributing to $\Pi(Q^2)$,
whereas those for $\PiT(Q^2)$ are replaced by the constant mass term.
This also means that the tadpole diagram drops out completely since it  contributes to $\PiT(Q^2)$ only.
For the same reason we also do not need to worry about quadratic divergences;
all self-energy diagrams are only logarithmically divergent and these divergences
are removed by the subtraction.

       \begin{figure*}
  \begin{center}
  \includegraphics[width=1\textwidth]{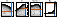}
  \end{center}
  \vspace{-5mm}
  \caption{UV exponents in Eq.~\eqref{uv-exp} and the maximum of the gluon dressing plotted over
           $c$. Each curve corresponds to a different value of $\alpha$ (same as in Fig.~\ref{fig:results-S2}).
           The orange lines are the physical values and the orange band for $Z_\text{max}$ is the lattice result.}\label{fig:UV-exponents}
  \end{figure*}

Fig.~\ref{fig:results-S2} shows the results for the running coupling $\conjg\alpha(x)$, the ghost dressing $G(x)$, the gluon dressing $Z(x)$ and the three-gluon vertex dressing $F_{3g}(x)$ in Setup 3, i.e.,  with a back-coupled three-gluon vertex and including the two-loop terms in the gluon DSE. The calculations were performed at a fixed value $\beta = 0.001$. The variation with $\beta$ will be discussed in Sec.~\ref{flow-lines}. The different curves correspond to a variation of the coupling parameter $\alpha$ over its full range. The running coupling is invariant under renormalization but the propagator and vertex dressing functions are not, so we plot the bare dressing functions according to Eq.~\eqref{ren-consts} for better visibility, i.e.,
\begin{equation}
	G^{(B)} = Z_c\,G\,, \quad
	Z^{(B)} = Z_A \,Z\,, \quad
	F_\text{3g}^{(B)} = \frac{F_\text{3g}}{Z_{3g}}\,.
\end{equation}
With increasing but finite $\alpha$, the ghost dressing function in Fig.~\ref{fig:results-S2} rises in the IR and saturates at a finite value. The gluon dressing behaves like $Z(x) \propto x$ in the IR and develops a maximum at intermediate momenta. As a consequence, also the running coupling $\conjg\alpha(x) = Z(x)\,G(x)^2$ vanishes like $Z(x) \propto x$ in the IR and develops a maximum, however at a different value of $x$. The three-gluon vertex becomes increasingly smaller in the IR, where it eventually crosses zero and diverges logarithmically. The orange bands in Fig.~\ref{fig:results-S2} mark the onset of the `QCD-like' decoupling solutions around $\alpha \lesssim 1$, as we discuss below.

For $\alpha\to \infty$, the curves eventually converge to the scaling solution, where the dressing functions in the IR behave like
\begin{equation}
	G(x) \propto x^{-\kappa}\,, \quad
	Z(x) \propto x^{2\kappa}\,, \quad
	F_{3g}(x) \propto x^{-3\kappa}\,.
\end{equation}
In this way, the scaling solution is the envelope of the decoupling solutions
and produces a finite running coupling $\conjg\alpha(x)$ which becomes constant in the IR.
Thus, the emergence of a \textit{family} of decoupling solutions is simply a consequence of varying the coupling parameter $\alpha$,
and the scaling solution is the limiting value for $\alpha\to\infty$.

The coupled DSEs provide us with an internal way to quantify the truncation error.
In practice the equations do not converge unless we modify the renormalization constant $Z_{3g}$ in Eq.~\eqref{ren-constants-2} by a parameter $c$,
\begin{equation}\label{def-c}
	Z_{3g} = c\,\frac{Z_A}{Z_c}  \qquad  \text{with} \qquad 0 < c < 1\,.
\end{equation}
This is equivalent to an additional renormalization of the three-gluon vertex in Eq.~\eqref{dses-1} of the form
\begin{equation}
	\begin{split}
		F_{3g}(Q^2) &= F_\mu + \mM(Q^2) - \mM(\mu^2)\,, \\[1ex]
		Z_{3g} &= F_\mu - \mM(\mu^2) \neq \frac{Z_A}{Z_c}\,.
	\end{split}
\end{equation}
Note that this additional renormalization would not be necessary if the STIs were preserved. Since we are working in different truncations defined by the Setups 1, 2 and 3, the STIs are no longer satisfied, but the effect can be compensated by introducing the parameter $c \neq 1$. Note however that the STIs constitute a further system of functional relations, and in non-trivial nonperturbative approximations such as the present truncations of the vertex expansions it is impossible to satisfy all sets of functional relations simultaneously;
see e.g.~\cite{Fischer:2008uz, Cyrol:2016tym, Cyrol:2017ewj, Dupuis:2020fhh} for respective discussions.
From Eqs.~(\ref{running-coupling-0}--\ref{running-couplings-0}),
the parameter $c^2$ translates to the ratio $\conjg\alpha_{3g}(x)/\conjg\alpha(x)$ evaluated in the UV.
This reflects the consistency of the running couplings in the UV, which is known to be  important for the quantitative accuracy of the solutions.

It turns out that for each truncation there is a value $c_\text{max}$
where the anomalous dimensions of the ghost and gluon propagators reach their physical values. In each iteration step we employ fits of the form
\begin{equation}\label{uv-exp}
	G(x\to \infty) = \frac{a_\text{gh}}{(\ln x)^{\gamma_\text{gh}}}\,, \quad
	Z(x\to \infty) = \frac{a_\text{gl}}{(\ln x)^{\gamma_\text{gl}}}\,,
\end{equation}
where the coefficients and exponents are free fit parameters. Thus, they are not forced to reproduce the anomalous dimensions of Yang-Mills theory given by
\begin{equation}\label{anomalous-dim}
	\gamma_\text{gh} = \frac{9}{44}\,, \qquad
	\gamma_\text{gl} = \frac{13}{22}\,, \qquad
	2 \gamma_\text{gh} + \gamma_\text{gl}= 1\,.
\end{equation}
Fig.~\ref{fig:UV-exponents} shows the resulting UV powers, plotted as a function of $c$. The $\beta$ value is fixed, and we scan the full $\alpha$ range. For $\alpha\to 0$, the anomalous dimensions vanish and for $\alpha\to\infty$ they converge to the scaling solution. The largest $c$ value where we obtain convergent solutions is marked by the vertical dashed lines. To the right of these lines we extrapolated each curve by a cubic fit with free fit parameters. The orange lines denote the physical values in Eq.~\eqref{anomalous-dim}. One can see that above a certain value of $\alpha$, which marks the onset of the decoupling solutions, the curves show a substantial curvature and extrapolate to their physical values approximately at the same point $c_\text{max} \approx 0.96 \dots 0.97$. This suggests to identify $c_\text{max}$ with the physical point in the truncation.

The rightmost plot in Fig.~\ref{fig:UV-exponents} shows the maximum value of the gluon dressing function $Z_\text{max} = \max Z^{(B)}(x)$ as a function of $c$. Close to $c_\text{max}$, there is a strong curvature which even appears to tend to infinity for the scaling solution. The horizontal orange band corresponds to the region of lattice results for $Z_\text{max}$, see below. The extrapolated values are compatible with this band, although the extrapolation induces a large uncertainty.

\begin{figure*}
	\begin{center}
		\includegraphics[width=0.95\textwidth]{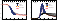}
	\end{center}
	\caption{Comparison of our DSE results for the ghost and gluon dressing functions with lattice data from Ref.~\cite{Duarte:2016iko}.
		The solid black curves are the DSE results (for $Z(x)$ we plot the range in $c \in [0.5,0.95]$ for illustration)
		and the orange dashed curves are the lattice parametrizations from Eq.~\eqref{lattice-params}.
		The red curves for $Z(x)$ interpolate between the lattice data and the best DSE result; 
		the corresponding curves in the left panel are obtained by solving the ghost DSE using this input.
	}\label{fig:lattice-comparison}
\end{figure*}

The results in Setups 1 and 2 are qualitatively similar to those in Fig.~\ref{fig:UV-exponents}, except with a different $c_\text{max}$:
\begin{itemize} \renewcommand{\itemsep}{0.1pt}
	\item Setup 1: $c_\text{max} \approx 0.59 \dots 0.61$,
	\item Setup 2: $c_\text{max} \sim 0.89 \dots 0.90$,
	\item Setup 3: $c_\text{max} \sim 0.96 \dots 0.97$.
\end{itemize}
Without any truncation one must have $c = 1$, so the  value of $c_\text{max}$ allows us to quantify the truncation error. With a tree-level three-gluon vertex it is about $40\%$, which reduces to $\sim10\%$ if the three-gluon vertex is back-coupled and $3-4\%$ if also the two-loop terms are included. With further systematic improvements such as a back-coupling of the ghost-gluon and four-gluon vertices, the error should thus become even smaller. Indeed, then the respective $c$, or rather the ratio of all different couplings, approaches unity and the deviation is at the sub-percent level, see \cite{Cyrol:2016tym, Huber:2020keu}.

Close to the respective value $c_\text{max}$, the dressing functions in Fig.~\ref{fig:results-S2} are very similar in all setups. In double-logarithmic plots like those for $\conjg\alpha(x)$, $G(x)$ and $F_{3g}(x)$ (recall that asinh\,$F$ grows logarithmically for large $|F|$) they are hardly distinguishable, except that in Setup~1 the three-gluon vertex stays at tree-level. The quantity that is most sensitive to the truncations is the bump of the gluon dressing function whose height defines $Z_\text{max}$. In Setup~2 the bump is slightly larger and narrower than in Setup~3 because the two-loop terms in the gluon self-energy are positive (cf.~Fig.~\ref{fig-Zinv}) and therefore flatten $Z(x)$.

In any case, these observations imply that the gluon dressing function may well have only reached a fraction of its true height. The largest possible value $c=0.95$ we can reach in Setup~3 corresponds to the results in Fig.~\ref{fig:results-S2}. The same observation holds true, although less pronounced, for the ghost dressing at intermediate momenta. This makes it difficult to compare with lattice results since we should match the results at $c=c_\text{max}$ but the functions can change substantially in its vicinity. In addition, the lattice data are only determined up to scale setting (which amounts to horizontal shifts of the curves on a logarithmic scale) and multiplicative renormalization. Finally, we must identify the value of $\alpha$ that should be compared to the lattice-type decoupling solution.

In practice we construct simple parametrizations for the lattice data in Ref.~\cite{Duarte:2016iko},
\begin{equation}\label{lattice-params}
	\begin{split}
		G_\text{lat}(x) &= \frac{1}{1 + 1.9 \,x^{0.7} + 2 \,x^2} + \frac{0.38\, (1-e^{-x})}{\ln(1+x)^{\gamma_\text{gh}}}\,, \\[1ex]
		Z_\text{lat}(x) &= \frac{6.3\,x}{(1+x)^{2.2}} + 6x\,e^{-2.1x}   \\[1ex]
		&+ \frac{x^2}{(0.3+x)(0.6+x)}\,\frac{3.1\,(1-e^{-3x})}{\ln(1+x)^{\gamma_\text{gl}}} \,.
	\end{split}
\end{equation}
The fits in \eqref{lattice-params} implement the one-loop running with $\gamma_\text{gh}$, $\gamma_\text{gl}$ from Eq.~\eqref{anomalous-dim},  together with $G_\text{lat}(0) = 1$ and $Z_\text{lat}(x\to 0) \propto x$. They are determined up to normalization and rescaling and must be matched to the UV running of the DSE results. The behavior of the UV coefficient $a_\text{gh}$ in Eq.~\eqref{uv-exp} is analogous to Fig.~\ref{fig:UV-exponents}. The curves converge approximately to the same value at $c_\text{max}$ independently of the coupling $\alpha$, from where we can match the ghost dressing with the lattice. Because this changes $G_\text{lat}(0)$ and our $G(x)$ satisfies $G(0) = \sqrt{\alpha}$, with this we identify $\alpha \sim 2.5$ as the value corresponding to the lattice decoupling solutions. Matching the UV running of the gluon together with the condition $Z(1/\beta)=1$ then fixes the gluon dressing function and the scale.

From the resulting curves in Fig.~\ref{fig:lattice-comparison} one can see that there is a substantial gap between the lattice data and the DSE results obtained with $c=0.95$. However, we find a similar behavior when we apply naive cubic extrapolations like those in Fig.~\ref{fig:UV-exponents} to $G(x)$ and $Z(x)$ over the whole range in $x$: The extrapolated ghost dressing at $c=0.97$ reproduces the lattice data, and the extrapolated gluon dressing matches the height of the lattice data but its bump is somewhat shifted.

To check the self-consistency of our comparison, we also solved the ghost DSE for $G(x)$ in Eq.~\eqref{dses-redef-2}, with $\Sigma(x)$ from Eq.~\eqref{ghost-self-energy-redeff}, using a fixed gluon input. In this case the standalone ghost DSE converges without problems, since the gluon propagator is just an input function. For that purpose, we generated a family of curves for $Z(x)$ which interpolate between the DSE and lattice results (shown in the right panel of Fig.~\ref{fig:lattice-comparison}). The corresponding results for $G(x)$  from the ghost DSE are shown in the left panel. One can see that the DSE solution with the lattice gluon input reproduces the lattice ghost without difficulties. The ghost DSE only depends on $\alpha$. Solving it for different values of $\alpha$, we find that the quantity $\sum_x |G(x) - G_\text{lat}(x)|$ is minimized for $\alpha \approx 2.5$ in agreement with our findings above. This also confirms that the only missing ingredient of the ghost DSE, the dressing of the ghost-gluon vertex, can only have a minor effect.

The problem of not being able to reach the physical value $c_\text{max}$ in the present truncation appears to be related to the back-coupling of the three-gluon vertex
and, by extension, to the four-gluon vertex which appears in its respective DSE. Our results are stable when changing the number of grid points or using different grids, but  achieving convergence close to $c_\text{max}$ becomes increasingly difficult even with relaxation and Newton methods (cf.~App.~\ref{sec:numerics}).
We easily obtain large and narrow bumps for $Z(x)$ when we employ model ans\"atze for $F_{3g}(x)$, but this comes at the expense of additional parameters and scales. This would obscure the discussion of the results in view of the gluon mass gap, and we refrained from doing so. We can also reduce the gap between the lattice and DSE results for $Z(x)$ by identifying the lattice decoupling solution with DSE solutions at different values of $\alpha$, but in those cases we can no longer simultaneously describe the ghost dressing function.

Concerning the four-gluon vertex ansatz, we employed $F_{4g}(x) = G(x)^2/Z(x)$ but when we use instead $F_{4g}(x) = Z_{4g} = const.$ the results are very similar. However, in the vicinity of $c_\text{max}$ the shape of $F_{4g}(x)$ especially in the mid-momentum region appears to play an increasingly prominent role in (de-)stabilizing the solutions. In Fig.~\ref{fig-ym} one can see that the four-gluon vertex appears in the gluon DSE only in the sunset diagram, which turns out to be negligibly small, so it mainly enters indirectly through the swordfish diagram in the three-gluon vertex. The general four-gluon vertex involves many tensors, including three momentum-independent ones which may lead to additional effects~\cite{Eichmann:2015nra,Huber:2020keu}. For functional computations aiming at quantitative precision in Yang-Mills theory, which also implement  a dynamically back-coupled four-gluon vertex, we refer to Refs.~\cite{Cyrol:2016tym, Huber:2020keu}; e.g. in~\cite{Huber:2020keu} the deviation from the STI was quantified to be on the sub-percent level.

In summary, our analysis implies that the lattice results can be matched to the DSE results for a particular value of the coupling parameter $\alpha$. In the next section we will see that this generalizes to a particular line of constant physics in the $(\alpha,\beta)$ plane.
\begin{figure}[t]
	\includegraphics[width=1\columnwidth]{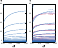}
	\caption{Lines of constant physics in the $(\alpha,\beta)$ plane.
		The left panel shows the results for Setup 3 and $c=0.92$ and
		the right panel those for $c=\{0.86,0.88,0.90,0.92\}$.}
	\label{fig-flow-lines}
\end{figure}

\subsection{Lines of constant physics}\label{flow-lines}

The results discussed so far were obtained at a fixed value of the mass parameter $\beta=0.001$. The remaining question in Scenario A is how the respective results and findings change with $\beta$. For modest changes with larger $\beta$, we find that the plots in Fig.~\ref{fig:results-S2} are essentially unchanged except they may be shifted horizontally in $\log x$. The overall scale is arbitrary  and we are free to rescale the solutions, and after rescaling they are almost identical. Thus, in this domain the mass parameter $\beta$ rescales the solutions but does not change any physics. This is in contradistinction to a variation of the coupling parameter $\alpha$, which  distinguishes physically different solutions.

\begin{figure}[t]
	\includegraphics[width=1\columnwidth]{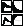}
	\caption{Dressing functions along the trajectories in Fig.~\ref{fig-flow-lines};
		each curve is a superposition of 30 curves for different $\beta$ in the interval $10^{-6} \leq \beta \leq 10^{-3}$.}
	\label{fig-flow-lines-results}
\end{figure}
For quantifying these observations, we show the lines of constant physics in the $(\alpha,\beta)$ plane in Fig.~\ref{fig-flow-lines}. For their determination
we extracted the maximum of the running coupling $\conjg\alpha(x)$ for each parameter set $(\alpha,\beta)$. Then, we rescaled $x$ such that the maximum always appears at the same point $x=x_0$. Accordingly, a given line is defined by the same value $\conjg\alpha(x_0)$. This amounts to interpolating the dressing functions at fixed value of $\beta=0.001$ and finding the value of $\alpha$ for a given $\conjg\alpha(x_0)$. Along each line, all functions are identical up to rescalings, which can be seen in Fig.~\ref{fig-flow-lines-results}:
every curve therein is a superposition of 30 curves along a given line and these curves are practically indistinguishable.

In Fig.~\ref{fig-flow-lines} one can see that for sufficiently large $\beta$ the lines of constant physics become indeed horizontal. On the other hand, if they were horizontal for \textit{all} values of $\beta$ this would be incompatible with physical expectations: In the massless theory with $\beta=0$, the coupling $\alpha$ should not produce physically different solutions but only rescale the system. In other words, in the massless limit the \textit{line} of constant physics should be the vertical axis in Fig.~\ref{fig-flow-lines}. These different forms can only be reconciled with each other if the lines bend towards the origin in the $(\alpha,\beta)$ plane. Indeed, this is what we find for very small values of $\beta$, as is visible in Fig.~\ref{fig-flow-lines}.

The behavior in Figs.~\ref{fig-flow-lines} and~\ref{fig-flow-lines-results} is completely analogous in Setups 1 and 2 and thus appears to be a general feature. In practice we cannot reach $\beta=0$ for the reasons discussed in Appendix~\ref{sec:numerics}. $1/\beta$ is the subtraction point in the gluon DSE, and the shape of the gluon self-energy  defines a calculable window for $\beta$. For increasing values of $\alpha$ or $c$, the gluon dressing becomes increasingly narrow and this window shrinks. The smallest value we can reach is typically $\beta \sim 10^{-6}$. The results in the left panel of Fig.~\ref{fig-flow-lines} were obtained with $c=0.92$; in the right panel one can see that the effect becomes more pronounced when $c$ is increased towards its practical limit $c=0.95$, i.e., the curves become steeper. However, for $c \gtrsim 0.92$ we can no longer cover the full range in $\beta$.

\begin{figure}[t]
	\includegraphics[width=0.55\columnwidth]{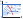}
	\caption{Sketch of the $(\alpha,\beta)$ plane.}
	\label{fig-alpha-beta-plane-2}
\end{figure}
The interpretation of our results is sketched in Fig.~\ref{fig-alpha-beta-plane-2}. The two parameters $(\alpha,\beta)$  in the Yang-Mills system form a combination $\lambda_0(\alpha,\beta)$ that acts along the lines of constant physics and only rescales the system without changing any physics. The second combination $\lambda_1(\alpha,\beta)$ is orthogonal to these lines and distinguishes the physically different solutions, i.e., only $\lambda_1$ is an active parameter. In other words, in the presence of a mass term the coupling parameter $\alpha$ no longer rescales the system but distinguishes the physically different decoupling solutions, with the scaling solution as their endpoint.

In the limit $\beta\to 0$, the shape of the lines implies that $\lambda_0(\alpha,0) = \alpha$, i.e., the coupling $\alpha$ only rescales the solutions while not changing the physics. However, this means that the scaling solution obtained at $\beta \neq 0$ and $\alpha\to\infty$ is identical to the scaling solution at $\beta=0$ with arbitrary $\alpha$, and it is the \textit{only} solution that remains in the massless limit. This implies that in the present Scenario A the only \textit{physical} solution is the scaling solution, as the mass parameter $\beta$ is merely an artifact that must be sent to zero. The scaling solution  is therefore a parameter-free, intrinsic property of the system.  Moreover, since the scaling solution as well as the whole family of decoupling solutions implement confinement via the gluon mass gap~\cite{Braun:2007bx, Fister:2013bh}, this implies confinement in the Landau gauge.

\begin{figure}[t]
	\includegraphics[width=1\columnwidth]{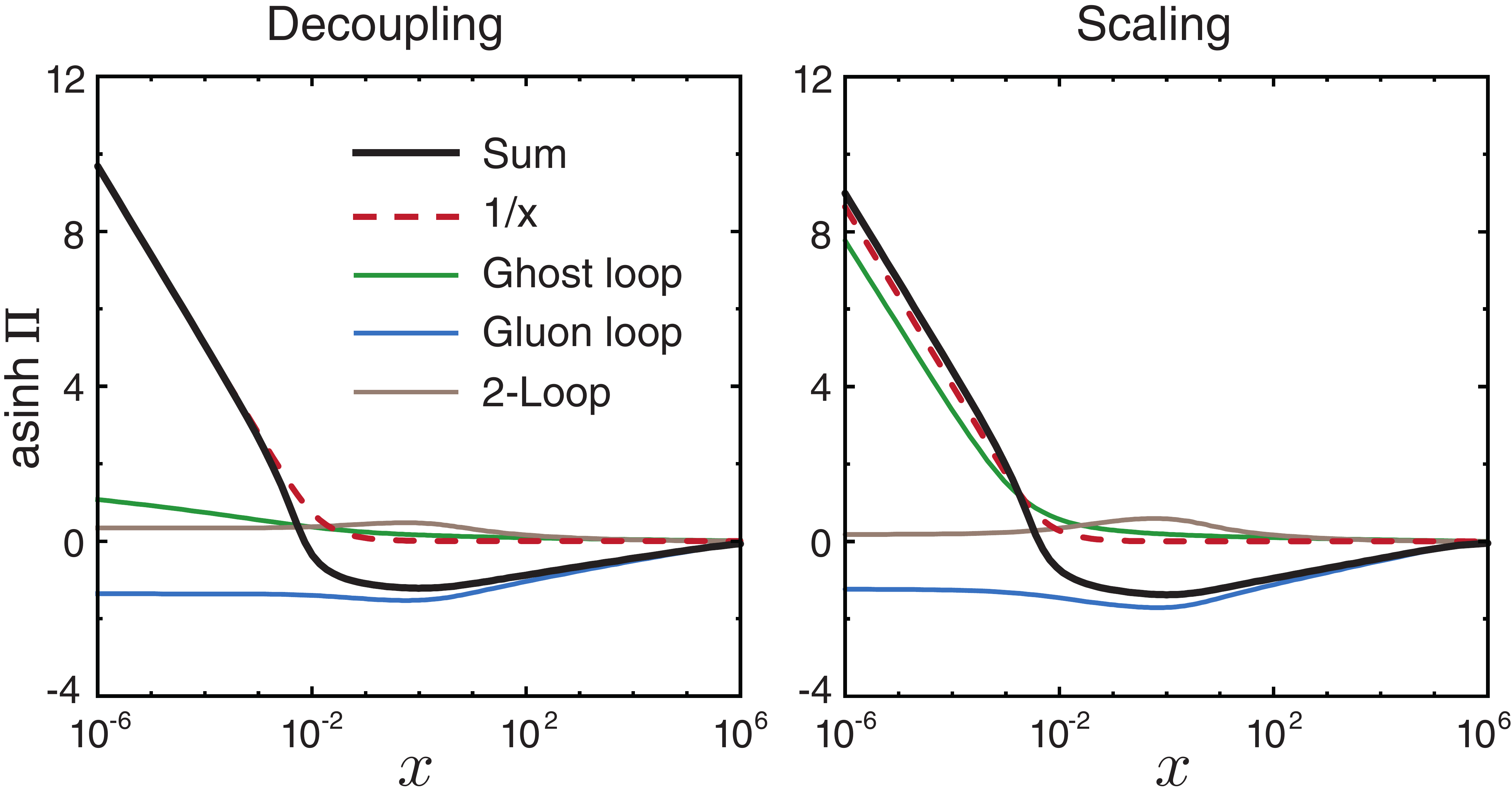}
	\caption{Contributions to the gluon self-energy $\mathbf{\Pi}(x)$ for the decoupling solution with $\alpha = 2.5$ and the scaling solution,
		both within Scenario A.}
	\label{fig-Zinv-A}
\end{figure}

While this seems theoretically appealing, there is a conceptual problem that can be traced back to the determination of the IR exponent $\kappa$ discussed around Eq.~\eqref{kappa-determination}. We have replaced $\PiT(Q^2)$ by a constant mass term of the form~\eqref{PiT-const}, which is `removed' in the end by sending $\beta\to 0$. As $\beta$ only appears in the term $\beta/x$, we could absorb it into the scale and thus sending $\beta\to 0$ is equivalent to renormalizing the equations at $x = 1/\beta \to \infty$. This entails, however, that the mass term $1/x$ cannot truly be removed as shown in Fig.~\ref{fig-Zinv-A}, which is the decomposition of the gluon self-energy $\mathbf\Pi(x)$ in Scenario A. We emphasize that its analogue in Fig.~\ref{fig-Zinv} corresponds to Scenario B, which we discuss below. For fixed $\beta$ and $\alpha\to\infty$ the ghost loop becomes large,  but even for the scaling solution at $\alpha\to\infty$ the term $1/x$ is present and dominates the IR behavior. Had we not absorbed $\beta$ into the scale, for $\beta\to 0$ the same would happen but all curves would be pushed to the left on a logarithmic scale, so that the rise with $1/x$ in the IR eventually drops out of the numerical grid and is shifted to $\log x \to -\infty$. Therefore, the effect of the mass term is never  truly switched off.

The same rescaling effect explains the apparent discrepancy between the solutions for $Z(x)$ at $\alpha \to 0$ from  Eqs.~\eqref{dses-redef-1} and~\eqref{dses-redef-2}, where the first set of equations entails $Z(x) \to 1$ and the other $Z(x) \to 1/(1 + 1/x - \beta)$. In the former case, the structure effects are also pushed to $\log x \to -\infty$, so the renormalized theory never becomes a free massless theory for $\alpha \to 0$.

Because the $1/x$ term drives the IR properties, in this case the scaling solution at $x\to 0$ with
\begin{equation}
	G(x) \propto x^{-\kappa}\,, \quad
	Z(x) \propto x^{2\kappa}\,, \quad
	F_{3g}(x) \propto x^{-3\kappa}
\end{equation}
has an IR exponent $\kappa = 1/2$. This is also what we found analytically in Fig.~\eqref{fig-ir-exponent} when sending the parameter $\lambda$ to zero. The ambiguity for $\lambda=0$ reappears in the numerical solution in the following way: Since the mass term in Eq.~\eqref{PiT-const} is arbitary as long as we send $\beta\to 0$ in the end, we are free to replace it with any other ansatz of the form
\begin{equation} \label{kappa-mod}
	\frac{\PiT(Q^2)}{Q^2} = \frac{g^2}{4\pi}\,G_0^2 \left( \frac{\beta\mu^2}{Q^2}\right)^{2\kappa} \,,
\end{equation}
where $\kappa$ is a free parameter, as long as we remove this term again in the end by sending $\beta\to 0$.
After the redefinitions from Sec.~\ref{sec:renormalization-main} this becomes
\begin{equation}\label{kappa-replace}
	\frac{\PiT(x)}{x} \to \frac{1}{x^{2\kappa}}\,,
\end{equation}
which replaces the $1/x$ term in the DSEs~\eqref{dses-redef-2}.
Its effect is again negligible in the mid- and large-momentum regions,
so it does not change the behavior of the solutions, but as long as $\kappa \geq 1/2$ it still dominates the IR and leads to a scaling solution with IR exponent $\kappa$. We checked this numerically and found that by dialling $\kappa$ we can indeed generate any scaling solution we want, where the lines of constant physics behave qualitatively in the same way as before.

In other words, in Scenario A the limit $\beta\to 0$ is not unique, which is a manifestation of the 0/0 problem in the analytic IR analysis and leads to a `family of scaling solutions' depending on the arbitrary value of the IR exponent $\kappa$. Because this ambiguity does not seem very appealing, in the next section we revisit the initial  problem and discuss a possible alternative scenario.

\section{Gluon mass gap: Scenario B} \label{sec:scenario-b}

The second interpretation, which we call Scenario B, is that $\Delta_0(Q^2)$
is not an artifact but indeed a dynamical feature of the equations. To study it, we return to Eq.~\eqref{self-energy-delta-2}  with both $\Delta_0(Q^2)$ and $\Delta_L(Q^2)$ in the system. Then the transverse projection to arrive at $\mathbf\Pi(x) $ is achieved with $\lambda=1$ in Eq.~\eqref{projectors}. This is the transverse system that is commonly solved in DSE and fRG studies. However, in view of the previous discussion, it begs a few questions which we investigate below: the appearance of quadratic divergences in Sec.~\ref{sec:quad-div} and, after a brief discussion of the solutions, the issue of gauge consistency in  Sec.~\ref{sec-long-sing-main}, i.e., the vanishing of $\tilde \Pi$.

\subsection{Quadratic divergences}\label{sec:quad-div}

The first and more practical issue is that a dynamical $\Delta_0$ term potentially comes with quadratic divergences. While $\Delta_0$ vanishes in perturbation theory using dimensional regularization, a hard cutoff interferes with gauge invariance so that $\Delta_0$ has an intrinsic `artifact' admixture that must be removed. Once the quadratic divergences are subtracted in a fully consistent way, the remainder (if it is non-zero) must be a dynamical, nonperturbative effect.

Different methods have been employed to remove quadratic divergences in the sum $\mathbf{\Pi} = \Delta_T + \Delta_0/Q^2$,
see~\cite{Huber:2014tva} for an overview. One is an explicit subtraction at the level of the integrands; in that case one analyzes their UV behavior and subtracts the terms producing the quadratic divergences. This has been employed in what we call Setup~1~\cite{Fischer:2002eq} but it becomes cumbersome when the three-gluon vertex is back-coupled dynamically or two-loop terms are included. Another method is to subtract the quadratic divergences numerically by fitting $\mathbf{\Pi}$ to $1/Q^2$ terms~\cite{Fischer:2005en}. A consistent combination of both is typically done in the fRG, where the consistency follows from the (trivial) RG consistency of the fRG setup~\cite{Cyrol:2016tym, Dupuis:2020fhh}.

Having individual access to $\Delta_T$ and $\Delta_0$, a simpler method is to
subtract $\Delta_0$ directly, i.e.
\begin{equation}
	{\Delta_0 (Q^2)}\to {\Delta_0(Q^2) - \Delta_0(Q_0^2) + \frac{g^2}{4\pi}\, G_0^2\,\beta\mu^2}\,.
\end{equation}
Since $\Delta_0(Q_0^2)$ is a constant, the arbitrariness in the choice of $Q_0^2$
is absorbed by adding another arbitrary constant
proportional to $\beta\mu^2$, again with a prefactor that ensures the correct renormalization.
This leads to a similar form of the equations as in Scenario A, where we added a constant mass term with
mass parameter $\beta$. After the redefinitions leading to Eq.~\eqref{dses-redef-2}, the self-energy becomes
\begin{subequations}\label{eq:PiDyn}
\begin{equation}\label{Pi-Sc-B}
	\mathbf\Pi(x) = \Delta_T(x) + \frac{1+\sigma(x)}{x}\,,
\end{equation}
with $\sigma(x)$ given by
\begin{equation}	\label{eq:sigma}
	\sigma(x) =  \Delta_0(x)-\Delta_0(x_0) \,.
\end{equation}
\end{subequations}
This procedure is equivalent to the one in Ref.~\cite{Huber:2020keu}, where a term $\propto 1/x$ is subtracted from the self-energy.
In this way, the only difference between the Scenarios A and B is the term $\sigma(x)$.
Note that we can no longer change the power of the $1/x$ term like we did in Eq.~\eqref{kappa-replace} in Scenario A,
because the subtraction must be done in its numerator.
As a consequence, the IR exponent $\kappa$ is no longer arbitrary
but a dynamical result with a definite value depending on the truncation.
Because $\sigma(x)/x$ is only active in the IR,
$x_0$ should not be too large to ensure numerical stability; in practice we choose $x_0=1$.
The arbitrariness of $x_0$ is then absorbed in the change of $\beta$ which determines the subtraction point $x=1/\beta$.

\begin{figure}[t]
	\includegraphics[width=1\columnwidth]{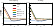}
	\caption{Contributions to the gluon self-energy $\mathbf{\Pi}(x)$ for the decoupling solution with $\alpha = 2.5$ and the scaling solution,
		both within Scenario B with the extra term $\sigma(x)$.}
	\label{fig-Zinv-B}
\end{figure}

\subsection{Discussion of solutions}

The numerical results in Scenario B do not require a separate discussion, because
the only change in the equations is the addition of the term $\sigma(x)$ in Eq.~\eqref{Pi-Sc-B},
which only affects the IR region. Thus, in the mid- and high-momentum
regions, and also at low momenta as long as $\alpha$ is not too large, all previous statements remain intact.
The shape of the solutions in Fig.~\ref{fig:results-S2}, the behavior of the UV exponents in Fig.~\ref{fig:UV-exponents},
and the values of $c_\text{max}$ for the different setups all remain approximately the same in Scenario B.
The only noticeable difference appears in the IR, where $\sigma(x)$ is active.

Fig.~\ref{fig-Zinv-B} shows the gluon self-energy contributions in this case.
These are the same plots as in Fig.~\ref{fig-Zinv} except now we separate the $1/x$ and $\sigma(x)/x$ contributions explicitly.
For the decoupling solutions, $\sigma(0)$ is a constant.
The remaining self-energy terms in $\Delta_T(x)$ are either constant or diverge at best logarithmically in the IR (see App.~\ref{app:expl}),
so that $\mathbf\Pi(x)$ and thus $Z(x)^{-1}$ are dominated by the terms $\propto 1/x$ for $x\to 0$.
These terms generate a mass gap such that the (dimensionless) gluon propagator at the origin reads
\begin{equation}
	D(0) = \frac{Z(x)}{x}\Big|_{x\to 0} =  \frac{1}{1 + \sigma(x)}\,.
\end{equation}
After reinstating dimensions through $x=Q^2/(\beta\mu^2)$, the r.h.s. is divided by
the (arbitrary) factor $\beta\mu^2$,
from where one can extract the gluon mass scale according to Eq.~\eqref{gluon-prop-0}.

When approaching the scaling solution for $\alpha\to\infty$, $\sigma(x\to 0)$ grows and eventually diverges with
a power $x^{1-2\kappa}$ stemming from its ghost loop contribution.
At the same time the ghost loop contribution to $\Delta_T(x)$
also diverges with $x^{-2\kappa}$ in the scaling limit.
Taken together, both $\Delta_T(x)$ and $\sigma(x)/x$ contribute with the same power to the IR divergence of $\mathbf\Pi(x) \propto x^{-2\kappa}$,
which for $\kappa > 1/2$ beats the divergence from the mass term $1/x$
as is visible in the right panel of Fig.~\ref{fig-Zinv-B}.

\begin{figure}
	\begin{center}
		\includegraphics[width=1\columnwidth]{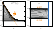}
	\end{center}
	\vspace{-5mm}
	\caption{Left: The infrared exponent $\kappa$ for the scaling solution
		can be read off from the envelope of the decoupling solutions.
		Right: Dependence of $\kappa$ extracted from Eq.~\eqref{ghost-kappa} on the parameters $c$ and $\alpha$.
	}\label{fig:scb-plots}
\end{figure}

As a result,
the dressing functions in the IR scale with
\begin{equation}
	G(x) \propto x^{-\kappa}\,, \quad
	Z(x) \propto x^{2\kappa}\,, \quad
	F_{3g}(x) \propto x^{-3\kappa}\,.
\end{equation}
The IR exponent $\kappa$ can be read off from any of the dressing functions in the scaling limit. This is shown in the left panel of Fig.~\ref{fig:scb-plots} for the ghost dressing, where the line $1/x^\kappa$ is the envelope of the decoupling solutions. In practice we determine $\kappa$ from the ghost self-energy by fitting it to $\Sigma(x) = a + b x^\kappa $ for $x\to 0$, where $a$, $b$ and $\kappa$ are free fit parameters depending on $\{\alpha,\beta,c\}$, i.e., also for the decoupling solutions. From Eq.~\eqref{dses-redef-2} we deduce that the ghost dressing function in the IR is well approximated by
\begin{equation}\label{ghost-kappa}
	G(x)^{-1} \stackrel{x\to 0}{\longlongrightarrow} \frac{1}{\sqrt{\alpha}} + b x^\kappa\,,
\end{equation}
so that scaling at $x\to 0$ is achieved for $\alpha\to\infty$ whereas for each decoupling solution $G(0) = const$.
For sufficiently large $\alpha$ we may read off the scaling exponent $\kappa \approx 0.58$, which is compatible with contemporary DSE~\cite{Huber:2018ned} and fRG results~\cite{Cyrol:2016tym}. Note however that the approximate value undershoots the analytic scaling $\kappa = 0.595$ that is obtained in the present approximation for $\alpha\to \infty$.
In the right panel of Fig.~\ref{fig:scb-plots} one can see that $\kappa$ is almost independent of $c$, i.e., the fact that we cannot exhaust the full range up to $c_\text{max} \approx 0.96 \dots 0.97$
is irrelevant for its determination.

\begin{figure}[t]
	\includegraphics[width=0.88\columnwidth]{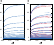}
	\caption{Lines of constant physics in the $(\alpha,\beta)$ plane for Scenario B and Setup 3,
		calculated for $c=0.90$ (left) and $0.86 \leq c \leq 0.90$ (right).}
	\label{fig-flow-lines-SCB}
\end{figure}

When we solve the DSEs in the $(\alpha,\beta)$ plane
we find the same behavior as in Scenario A.
One can identify lines of constant physics, which for $\beta \to 0$ bend towards the origin as shown in Fig.~\ref{fig-flow-lines-SCB}.
The interpretation is thus the same:
One combination of $\alpha$ and $\beta$ rescales the solutions and another one distinguishes the decoupling solutions,
with the scaling solution as their endpoint.

The crucial difference to Scenario A, however, is the fact
that $\beta$ is no longer an artificial mass parameter that must be removed in the end by sending $\beta\to 0$,
because now it arises from the subtraction of quadratic divergences.
Therefore, any value for $\beta$ is equally acceptable,
which also seems more appealing considering the conceptual problems with the limit $\beta=0$ encountered in Scenario A.
As a consequence, there is no longer a criterion that discriminates between the different solutions.
Instead of an ambiguity in the IR exponent $\kappa$, one is left with the ambiguity
between the scaling and decoupling solutions, which for $\beta \neq 0$ are distinguished by the parameter $\alpha$.
This raises the question: Can one find \textit{another} criterion that allows for a selection of solutions?

\subsection{Longitudinal singularities} \label{sec-long-sing-main}

The fundamental question in Scenario B is \textit{how} $\tilde \Pi(Q^2)$ can vanish, given that
the dynamical term $\Delta_0(Q^2) \neq 0$ now  contributes to it.
As we already mentioned, this implies a non-vanishing $\Delta_L(Q^2)$ for the consistency relation
$\Delta_L = -\Delta_0/Q^2$ to hold.
To this end, let us re-examine the assumptions we made. The crucial (implicit) assumption in our approximation so far is the use of the ghost-gluon vertex at tree-level. When introducing Scenario B in Section~\ref{sec:dses}, we already emphasized that the ghost-gluon vertex has two tensor structures, cf.~Eq.~\eqref{ggl-vertex-0}:
\begin{equation}\label{ggl-vertex}
	\Gamma^\mu_\text{gh}(p,Q) = -ig  f_{abc}\left[ (1+A)\,p^\mu + B\,Q^\mu\right]\,.
\end{equation}
The momentum $p^\mu$ is the outgoing ghost momentum and $Q^\mu$ is the incoming gluon momentum.
The corresponding dressing functions are $A(p^2,p\cdot Q,Q^2)$ and $B(p^2,p\cdot Q, Q^2)$.
Accordingly, the tensor structure missing in the present approximation produces to a solely longitudinal
contribution to the inverse  gluon propagator which belongs to the $\Delta_L$ part.

The ghost-gluon vertex is UV-finite in Landau gauge and does not need to be renormalized.
The function $A$ is known to be small~\cite{Huber:2012kd,Huber:2020keu},
which is why a tree-level vertex is a good approximation.
However, little is known about the function $B$ apart from the limit $p^\mu = Q^\mu$ where the incoming ghost momentum vanishes,
in which case the vertex becomes bare~\cite{Taylor:1971ff} and thus one has $B(p^2,p^2,p^2) = -A(p^2,p^2,p^2)$.

The Lorentz tensor $Q^\mu$ attached to $B$ is purely longitudinal and vanishes whenever the gluon leg is contracted with a transverse gluon propagator in Landau gauge, i.e., for all internal gluon lines. As such, it does not contribute to the ghost self-energy in Fig.~\ref{fig-ym}, but it \textit{does} contribute to the ghost loop in the gluon DSE. Because the term is longitudinal, it drops out from the transverse projection $\mathbf{\Pi} = \Delta_T + \Delta_0/Q^2$ in Eq.~\eqref{self-energy-delta-2}. Hence, the full transverse dressing is not affected by the non-classical dressing $B$ in the ghost-gluon vertex (modulo small effects from $A\neq 0$). However, the longitudinal projection picks up an extra term and generalizes to
\begin{align}
	\PiT(Q^2) &= \Delta_0(Q^2) + Q^2 \,\Delta_L(Q^2)\,, \label{PiTilde-general}
\end{align}
where $\Delta_0$ is the sum of all $\PiT$-like self-energy contributions listed in Appendix~\ref{app:expl} (up to $A\neq 0$ effects).
The longitudinal dressing $\Delta_L$ can be worked out in analogy to Eq.~\eqref{ghost-loop}:
\begin{align}	\label{eq:DeltaL}
	\Delta_L(Q^2) &= -\frac{g^2 N_c}{2}\! \int_k \frac{G(k_+^2)\,G(k_-^2)}{k_+^2\,k_-^2}\,B(k_+^2,k_+\cdot Q,Q^2)\,,
\end{align}
where $k_\pm = k \pm Q/2$.
The resulting DSEs read
\begin{align}\nonumber
	Z(Q^2)^{-1} &= Z_A + \mathbf\Pi(Q^2) \,, \\[1ex]
	L(Q^2)^{-1} &= 1 + \xi\,\frac{\PiT(Q^2)}{Q^2}\,. \label{dse-gluon-l-2}
\end{align}

However, Eq.~\eqref{PiTilde-general} allows us to eliminate the completely longitudinal term $\PiT(Q^2)$ as required by the STIs. For $\PiT=0$ we must have
\begin{equation}\label{PiT-condition}
	\Delta_L(Q^2) \stackrel{!}{=} -\frac{\Delta_0(Q^2)}{Q^2}\,.
\end{equation}
For decoupling solutions, $\Delta_0(0)$ is a constant. Thus, Eq.~\eqref{PiT-condition} can only hold for all $Q^2$ if the longitudinal ghost-gluon dressing $B$ diverges like $1/Q^2$ for $Q^2\to 0$: the ghost-gluon vertex must exhibit a \textit{longitudinal massless pole}.
In this case $\widetilde{\Pi}$ vanishes, and we arrive at the purely transverse self energy
\begin{equation}
	\Pi^{\mu\nu}(Q) = \mathbf\Pi(Q^2)\left( Q^2\,\delta^{\mu\nu} - Q^\mu Q^\nu\right)
\end{equation}
with $\mathbf\Pi= \Delta_T+\Delta_0/Q^2$. The equation for $Z(Q^2)$ is unchanged and $B$ only serves to eliminate $\PiT(Q^2)$.

\subsubsection{Complete infrared dominance of the ghost}

Note that $\Delta_L(Q^2)$ in Eq.~\eqref{eq:DeltaL} only constitutes the part of the longitudinal dressing that originates in the ghost loop. However, while other diagrams in principle may also contribute to $\Delta_L(Q^2)$, the dominant contribution in the IR limit $Q^2 \to 0$ is given by \eqref{eq:DeltaL}. In particular, for the scaling solution all the other terms rise with lower powers of $1/Q^2$. In turn, this complete IR dominance of the ghost part for the scaling solution is only successively weakened for the family of decoupling solutions.

For completeness we also consider the sub-leading contributions from the gluon loops. In principle, a longitudinal tensor in the ghost-gluon vertex also propagates to the three-gluon vertex through the ghost triangle. However, we find that in the symmetric limit for the vertex such a term only gives a contribution to one out of three possible tensors. This contribution drops out from the gluon DSE because in the Landau gauge it is contracted with internal transverse gluon lines, see the discussion around Eqs.~\eqref{3gv-tensors} and~\eqref{ghost-triangle} in the appendix. Therefore, in the truncations we consider herein such a longitudinal term can only affect the ghost loop in the gluon self-energy, i.e., it can only arise from the ghost-gluon vertex. This argument also further corroborates the IR dominance of the ghost contribution in the gluon self energy for decoupling solutions that are sufficiently close to the scaling limit.
Based on this argument we proceed with the assumption of \textit{complete IR dominance} of the ghost contributions.

\begin{figure}[t]
	\includegraphics[width=1\columnwidth]{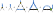}
	\caption{Ghost-gluon vertex DSE. The vertices where the $B$ term from Eq.~\eqref{ggl-vertex} survives are marked in yellow.}
	\label{fig-ggl-dse}
\end{figure}
If there are indeed longitudinal poles in
the ghost-gluon vertex, they should  dynamically emerge in its DSE which is shown in Fig.~\ref{fig-ggl-dse}.
While we cannot make any statement about the ghost-gluon four-point function in the last diagram therein (for a corresponding discussion, see Ref.~\cite{Alkofer:2011pe}),
it is clear that the longitudinal tensor must drop out for the internal vertices
since they are contracted with transverse gluon lines.
Thus, the $B$ term can only survive in the upper vertex in the second diagram (marked in yellow).
This yields an inhomogeneous Bethe-Salpeter equation (BSE) for $B$
with the structural form (see Appendix~\ref{sec:long-sing} for details)
\begin{equation}\label{iBSE}
	B = B_0 + \mK B \quad \Leftrightarrow \quad  B = (1-\mK)^{-1}\,B_0\,.
\end{equation}
Here, $\mK B$ represents the second (Abelian-like) diagram in the r.h.s. of Fig.~\ref{fig-ggl-dse} and the inhomogeneous term $B_0$ is the sum of the remaining loop diagrams; the tree-level term does not contribute to $B$.

\pagebreak

\begin{figure}[t]
	\includegraphics[width=0.8\columnwidth]{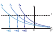}
	\caption{Typical eigenvalue spectrum of a BSE.}
	\label{fig-BSE-EV}
\end{figure}

Eq.~\eqref{iBSE} will produce a singularity in $B$ at some value $Q^2 = -m_i^2$ if the kernel satisfies $\mK = 1$, i.e. if any of its eigenvalues satisfy $\lambda_i(Q^2 = -m_i^2)=1$. In that case, the numerator defines a Bethe-Salpeter amplitude $\varphi$ via
\begin{equation}\label{bs-amp}
	B(p^2,p\cdot Q, Q^2)\, \stackrel{Q^2 \to -m_i^2}{\longlonglongrightarrow}\, \frac{\varphi_i(p^2,p\cdot Q,Q^2)}{Q^2 + m_i^2}\,.
\end{equation}
If one is only interested in the pole locations, one can equivalently solve the homogeneous BSE $\mK \varphi_i = \lambda_i \varphi_i$, which does not require knowledge of $B_0$ and depends on the Abelian-like diagram in Fig.~\ref{fig-ggl-dse} only.

The typical eigenvalue spectrum of a BSE kernel is sketched in Fig.~\ref{fig-BSE-EV}.
The eigenvalues $\lambda_i(Q^2)$ decrease with $Q^2$, which is also true in our case because the ghost propagators in the loop
depend on the external gluon momentum and fall off at large $Q^2$.
For $Q^2 > 0$, the eigenvalues must be smooth functions of $Q^2$ because the integrand does not have
spacelike singularities.
Each intersection with
$\lambda_i(Q^2 = -m_i^2) = 1$ corresponds to a singularity of $B$ at $Q^2 = -m_i^2$, which
may appear at timelike values $Q^2 < 0$. The largest eigenvalue $\lambda_0$
(the `ground state') corresponds to the  pole closest to the origin in $Q^2$.
In particular, if $\lambda_0$ satisfies $\lambda_0(Q^2 = 0) = 1$,
then $B$ must have a massless pole.

We can then solve the homogeneous BSE directly at $Q^\mu=0$ and calculate $\lambda_0(0)$.
Setting $A=0$ in Eq.~\eqref{ggl-vertex} to be consistent with our truncation of the DSEs,
the resulting equation becomes very simple,
\begin{equation}\label{hom-bse}
	\begin{split}
		&\int dx'\,\mK(x,x')\,\varphi(x') =  \lambda_0\,\varphi(x)\,, \\[1ex]
		& \mK(x,x') = \frac{N_c }{(2\pi)^2}\,x\,G(x')^2\,\int_{-1}^1 dy \,(1-y^2)^\frac{3}{2} \frac{Z(w)}{w^2}\,,
	\end{split}
\end{equation}
with $w = x+x'-2\sqrt{xx'}\,y$. Here we employed the same redefinitions and rescaling that led to Eq.~\eqref{dses-redef-2} for the coupled DSEs, i.e., $x$ is the dimensionless variable corresponding to $p^2$. The kernel only depends on the ghost and gluon dressing functions $G$ and $Z$.

\begin{figure*}[t]
	\includegraphics[width=0.9\textwidth]{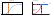}
	\caption{\textit{Left:}
		Largest eigenvalue of the homogeneous BSE~\eqref{hom-bse}. For the scaling solution $\alpha\to\infty$,
		the eigenvalue approaches 1 which corresponds to a longitudinal massless singularity in the ghost-gluon vertex. \textit{Right:} Sketch of the regime with a massless singularity (blue with $\lambda_0(Q^2=0)=1$) and the massive Yang-Mills regime without such a massless singularity (red with $\lambda_0(Q^2=0)<1$).}
	\label{fig-long-poles}
\end{figure*}

The left panel of Fig.~\ref{fig-long-poles} shows the largest eigenvalue $\lambda_0(Q^2=0)$ of $\mK$ as a function of the coupling parameter $\alpha$.
For $\alpha \to 0$, it is clear from Eq.~\eqref{hom-bse} that the eigenvalue must be zero because the product $ZG^2$ contains an intrinsic factor $\alpha$. On the other hand, the limiting value is $\lambda_0(0) =1$ because  $\lambda_0(0) > 1$ would produce a pole for spacelike values $Q^2>0$ and thus a tachyonic singularity. One can see in Fig.~\ref{fig-long-poles} that for all decoupling solutions with finite $\alpha$ one has $\lambda_0 \neq 1$. However, as $\alpha$ rises and approaches the scaling solution at $\alpha\to\infty$, the eigenvalue approaches $\lambda_0\to 1$. Thus, we find that the scaling solution indeed has a longitudinal massless singularity in the ghost-gluon vertex but the decoupling solutions do not. As a consequence, for the decoupling case the condition~\eqref{PiT-condition}, which ensures $\PiT(Q^2)=0$ and therefore gauge invariance, cannot hold. In conclusion, in Scenario B with the assumption of \textit{complete infrared dominance} of the ghost only the scaling solution survives.

The results in Fig.~\ref{fig-long-poles}  have been obtained within the Setup 3, but we find the same behavior in Setups 1 and 2. The result is also independent of the parameter $c$, which is consistent with the observation that the scaling properties are independent of $c$ (cf.~Fig.~\ref{fig:scb-plots}). In the analogous plot for Scenario A, $\lambda_0$ does not reach the value 1 but this is also not necessary because the condition~\eqref{PiT-condition} does not arise.

We also note that a necessary requirement for a massless singularity in the ghost-gluon vertex is the fact that the vertex in the present formulation is not completely ghost-antighost symmetric; see e.g.~\cite{Mader:2013eqo,Huber:2014tva} for more general discussions. If that were the case, $\mK B$ could not produce a massless pole strong enough to ensure Eq.~\eqref{PiT-condition} (see Appendix~\ref{sec:long-sing} for details). Another remark is that the ghost-gluon vertex DSE in Fig.~\ref{fig-ggl-dse} is the so-called `$c$ DSE', where the vertex on the left in each diagram is bare. An equivalent form is the `$A$ DSE', where all top vertices are bare (see e.g. Fig.~2 in~\cite{Huber:2018ned}). In the latter case one would not arrive at a BSE for the $B$ term and the longitudinal poles would need to come from higher $n$-point functions such as the four-ghost  or two-ghost-two-gluon vertices in the $t$ channel~\cite{Alkofer:2011pe}.

\subsubsection{General case}\label{eq:GeneralBSE}

The results of the last section beg the question whether the present setup or an (implicit) IR completion of the Landau gauge only admits the scaling solution. Indeed, the complete IR dominance of the ghost assumed above is only valid in the scaling limit for $\alpha\to \infty$. Accordingly, one should interpret the above result as a corroboration of a gauge-consistent scaling solution and not as an exclusion of gauge-consistent decoupling solutions. For the latter solutions there is no IR suppression of the gluonic contributions with powers of $Q^2$, and one may have to also include the three-gluon vertex in the analysis.

The coupled BSE system of ghost-gluon and three-gluon vertex has been studied intensively for lattice-type decoupling solutions in the PT-BFM scheme~\cite{Aguilar:2006gr, Binosi:2009qm, Aguilar:2011xe, Aguilar:2015bud, Aguilar:2016ock, Aguilar:2017dco,Aguilar:2021okw}. As discussed earlier in Sec.~\ref{ScA-solutions}, in our present setup the lattice solution roughly corresponds to $\alpha_\text{Lat}=2.5$ at $\beta=0.001$, which is shown by the vertical dashed line in the left of Fig.~\ref{fig-long-poles}. In particular, in \cite{Aguilar:2017dco}, within the approximations used there, an almost complete dominance of the gluonic contributions has been found. This may entail that for $\alpha_\text{Lat}$ the BSE for the ghost-gluon vertex is only fully consistent if also including the three-gluon term in Fig.~\ref{fig-ggl-dse}. Then, the BSE in \eqref{iBSE} generalizes to
\begin{equation}\label{iBSEgen}
	B = \left[1-\left(\mK_\textrm{gh-gl}+\mK_\textrm{3gl}\right)\right]^{-1}\,B_0\,,
\end{equation}
where $\mK_\textrm{gh-gl}$ is the contribution from the first diagram in Fig.~\ref{fig-ggl-dse} that we considered so far. In turn, $\mK_\textrm{3gl}$ stands for the second diagram in Fig.~\ref{fig-ggl-dse} with the three-gluon vertex. The ghost-gluon BSE in \eqref{iBSEgen} has to be augmented with that for the three-gluon vertex and has to be solved simultaneously for general $\alpha$.

While being of eminent importance, a detailed analysis is deferred to future work. Here we simply put forward a possible scenario that encompasses the present consistent scaling solution with complete IR ghost dominance and the lattice-type decoupling solution in \cite{Aguilar:2017dco} with relative three-gluon dominance. In combination, the present findings and that in \cite{Aguilar:2017dco} suggest that the massless singularity is present in the system for $\alpha\gtrsim \alpha_\text{Lat}$. Moreover, within a crude linear analysis the value of $\lambda_0(Q^2=0)$ for the lattice-type decoupling solution with $\alpha\approx \alpha_\text{Lat}$, cf. left panel in Fig.~\ref{fig-long-poles}, gives a rather small estimate for the contribution of $\mK_\textrm{gh-gl}$ to the required full eigenvalue  $\lambda_0(Q^2=0)=1$, which suggests the dominance of the gluonic contribution. Such a scenario is fully compatible with the findings in \cite{Aguilar:2017dco}.

In turn, in massive Yang-Mills theory with a large explicit mass $m_0 \approx m_\textrm{gap} \to \infty$ we do not expect such a massless singularity to be present.
Moreover, as discussed earlier, $\alpha_\text{Lat}$ corresponding to the lattice solution is slightly above the onset of the decoupling solutions around $\alpha_0 \sim 1$ highlighted in Figs.~\ref{fig:results-S2} and~\ref{fig:UV-exponents} and thus the massive Yang-Mills regime is defined by $\alpha \lesssim \alpha_0$.
The same statement within a quantitative approximation has been shown in \cite{Cyrol:2016tym}.

In combination, this makes us speculate about the exciting possibility to distinguish the possible QCD-type solutions with $\alpha\gtrsim \alpha_0$ from that in massive Yang-Mills theory with $\alpha\lesssim \alpha_0$ by the existence or absence of the massless pole in the longitudinal sector. This presence of this singularity is indicated by
\begin{align}
	\lambda_0(Q^2=0)=1\,,
\end{align}
and the scenario described above is sketched in the right panel in Fig.~\ref{fig-long-poles}. Whether it can be validated within more sophisticated truncations remains to be seen.

\subsubsection{Other scenarios}

It has  also been speculated that  $\PiT(Q^2)$ could be nonzero because the longitudinal DSE in \eqref{dse-gluon-l-2} depends on the combination $\xi \,\widetilde \Pi(Q^2)/Q^2$. Since $\xi=0$ in Landau gauge, one would obtain $L(Q^2)=1$ even if $\PiT(Q^2) \neq 0$. However, even if $\PiT$ were non-zero in Landau gauge, $L=1$ must still hold in all other linear covariant gauges and thus   $\PiT$ would still need to vanish for $\xi \neq 0$, in which case it is a non-analytic function of the gauge parameter $\xi$.

This scenario is in direct contradiction with the STI: While it is seemingly natural in terms of the gluon propagator, the STI requires the gluon self-energy~\eqref{vac-pol} to be transverse, and hence $\PiT = 0$.

\section{Summary and conclusions} \label{sec:summary}

In this work we studied the coupled Dyson-Schwinger equations (DSEs) for the ghost propagator,
gluon propagator and three-gluon vertex in the Yang-Mills sector of QCD in Landau gauge using different truncations. We addressed several questions related to the origin of mass generation.

We clarified the role of renormalization and showed that the parameter that distinguishes the decoupling solutions is the coupling $\alpha$, which no longer rescales the system in the presence of a mass term. Such a dynamical mass term arises from the longitudinal projection of the gluon self-energy, which we called $\PiT(Q^2)$ and which in principle appears to break gauge invariance. To this end, we offered two scenarios to remedy the problem.

In Scenario A, we considered the mass term to be an artifact of the hard cutoff and/or truncation, so we replaced it by a constant mass parameter $\beta$ and studied the limit $\beta\to 0$. We investigated the lines of constant physics in the $(\alpha,\beta)$ plane and found that one combination of $\alpha$ and $\beta$ rescales the solutions while the other distinguishes the physically different decoupling solutions, with the scaling solution as their endpoint for $\beta\to 0$. While the limit $\beta\to 0$ cannot be reached in practice,  we nonetheless conclude that in this case only the scaling solution survives, however with an ambiguity in the infrared scaling exponent $\kappa$.

In Scenario B, we did not modify the DSEs by hand and the dynamical mass term remains in the system. As a consequence, one needs to subtract quadratic divergences which introduces again an arbitrary  mass parameter $\beta$. However, in this case the longitudinal projection of the gluon self-energy can only vanish if either of the vertices that enter in the gluon DSE has longitudinal massless poles. In the three truncations we studied, this leaves the ghost-gluon vertex as the only candidate. Using the assumption of \textit{complete} infrared dominance of the ghost, we found that the vertex  indeed has such a pole, which is only present for the scaling solution.

Indeed, this solution is the only one which exhibits complete infrared dominance of the ghost. Interestingly, the massless pole is not present within this complete ghost dominance setup for any decoupling solutions. However, for lattice-like decoupling solutions massless poles have been found within the PT-BFM scheme. In contradistinction to the scaling case, they are dominated by the gluonic contributions. The combined observations allowed us to put forward a scenario that encompasses the present consistent scaling solution with complete infrared ghost dominance and the lattice-type decoupling solution with three-gluon vertex dominance, for more details see Section~\ref{eq:GeneralBSE}. A detailed numerical study of this scenario will be presented elsewhere.

\bigskip
\textbf{Acknowledgments.}
We are grateful to Reinhard Alkofer, Christian Fischer, Markus Huber, Axel Maas and Joannis Papavassiliou for discussions
and a critical reading of the manuscript. This work was supported by the FCT Investigator Grant IF/00898/2015, the Advance Computing Grant CPCA/A0/7291/2020, by EMMI, and by the BMBF grant 05P18VHFCA. It is part of and supported by the DFG Collaborative Research Centre SFB 1225 (ISOQUANT) and the DFG under Germany's Excellence Strategy EXC - 2181/1 - 390900948 (the Heidelberg Excellence Cluster STRUCTURES).

 \begin{figure*}[t]
  \includegraphics[width=0.8\textwidth]{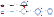}
  \caption{Momentum routing in the ghost and gluon DSEs (see Fig.~\ref{fig:2-loop} for the two-loop terms)}
  \label{fig-dses-mom-routing-1}
  \end{figure*}

 \newpage

 \appendix

\section{Explicit form of the DSEs}\label{app:expl}

  In this appendix we collect the expressions for the ghost propagator, gluon propagator and three-gluon vertex DSEs.
The propagator DSEs in Fig.~\ref{fig-dses-mom-routing-1} read
\begin{align}
	D_G^{-1}(Q) &= D_{G,0}^{-1}(Q) + \Sigma_G(Q)\,, \label{dse-ghost-0}\\
	(D^{-1})^{\mu\nu}(Q) &= (D_0^{-1})^{\mu\nu}(Q) + \Pi^{\mu\nu}(Q)\,, \label{dse-gluon-0}
\end{align}
where the full propagators depend on the ghost and gluon dressing functions $G(Q^2)$ and $Z(Q^2)$:
\begin{align}
	D_G(Q) &= -\frac{G(Q^2)}{Q^2}\,, \label{ghost-app}\\
	D^{\mu\nu}(Q) &= \frac{1}{Q^2}\left( Z(Q^2)\,T^{\mu\nu}_Q + \xi\,L^{\mu\nu}_Q\right)  \label{gluon-app}
\end{align}
Their tree-level expressions (with subscript 0) follow from the replacements
$G(Q^2)\to 1/Z_c$ and $Z(Q^2) \to 1/Z_A$, where $Z_c$ and $Z_A$ are the
ghost and gluon renormalization constants, respectively.
The transverse and longitudinal projectors are given by
$T^{\mu\nu}_Q = \delta^{\mu\nu} - Q^\mu Q^\nu/Q^2$ and $L^{\mu\nu}_Q = Q^\mu Q^\nu/Q^2$ and
$\xi$ is the gauge parameter.
Writing the ghost self-energy as $\Sigma_G(Q) = -Q^2\,\Sigma(Q^2)$ and decomposing the gluon self-energy as in Eq.~\eqref{vac-pol},
the DSEs take the form
\begin{align}
	G(Q^2)^{-1} &= Z_c + \Sigma(Q^2)\,, \label{ghost-dse-app} \\
	Z(Q^2)^{-1} &= Z_A + \Pi(Q^2) + \frac{\widetilde{\Pi}(Q^2)}{Q^2}\,,  \label{gluon-dse-t-app}
\end{align}
where the functions $\Pi(Q^2)$ and $\PiT(Q^2)$ are obtained from Eqs.~(\ref{projector-general}--\ref{projectors}) using $\lambda=0$.
According to the decomposition~\eqref{vac-pol}, we will refer to $\Pi(Q^2)$ as the \textit{transverse} part and $\PiT(Q^2)$ as the \textit{gauge} part in what follows.
Note however that for $\PiT(Q^2) \neq 0$ the transverse part $\Pi$ differs from the transverse \textit{projection} $\mathbf\Pi = \Pi + \PiT/Q^2$.

\begin{figure}[b]
	\includegraphics[width=0.9\columnwidth]{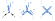}
	\caption{Kinematics for the ghost-gluon vertex, three-gluon vertex and four-gluon vertex}
	\label{fig-feynman}
\end{figure}

Every self-energy diagram contains a tree-level vertex, whose Feynman rules are collected in Fig.~\ref{fig-feynman} and Table~\ref{tab:tree-level}.
The full three-gluon vertex depends on 14 tensors~\cite{Ball:1980ax};
here we restrict ourselves to the classical tensor in Eq.~\eqref{feynman-rule-2}, which amounts to the replacement
\begin{equation}
	Z_{3g} \to F_{3g}(p_1^2,p_2^2,p_3^2)\,.
\end{equation}
In addition, we assume that the dressing function only depends on the symmetric variable:
\begin{equation}\label{3gv-approx}
	F_{3g}(p_1^2,p_2^2,p_3^2) \approx F_{3g}\left( \frac{p_1^2+p_2^2+p_3^2}{3}\right).
\end{equation}

Concerning the four-gluon vertex, it is convenient to rewrite the tree-level expression~\eqref{feynman-rule-3} as
\begin{equation}\label{4gv-tree-doublet}
	\Gamma^{\mu\nu\rho\sigma}_{4g,0}(p_1,p_2,p_3,p_4) =
	- g^2  Z_{4g}\, \sum_{i=1}^2 (\tau_i)_{ab,cd}\,\Gamma_i^{\mu\nu\rho\sigma}\,,
\end{equation}
where the  Lorentz and color tensors are given by
\begin{equation}
	\begin{split}
		\Gamma_1^{\mu\nu\rho\sigma} &= 2\delta^{\mu\rho}\,\delta^{\nu\sigma} - \delta^{\nu\rho}\,\delta^{\mu\sigma} - \delta^{\mu\nu}\,\delta^{\rho\sigma}\,, \\
		\Gamma_2^{\mu\nu\rho\sigma} &= 2\delta^{\mu\nu}\,\delta^{\rho\sigma} - \delta^{\nu\rho}\,\delta^{\mu\sigma} - \delta^{\mu\rho}\,\delta^{\nu\sigma}\,, \\
		(\tau_1)_{ab,cd} &= f_{abe}\,f_{cde}\,, \\
		(\tau_2)_{ab,cd} &= f_{ace}\,f_{bde}\,.
	\end{split}
\end{equation}
Because $(\tau_3)_{ab,cd} = f_{ade}\,f_{cbe}$ is linearly dependent due to the Jacobi identity ($\tau_3=\tau_1-\tau_2$),
Eqs.~\eqref{4gv-tree-doublet} and~\eqref{feynman-rule-3} are identical.

\begin{table*}[t]
	
	\normalsize

	\begin{align}
		\Gamma^\mu_\text{gh,0}(p,Q) &= -ig \,f_{abc}\,\widetilde{Z}_\Gamma\,p^\mu\,, \label{feynman-rule-1}\\
		ig\,f_{abc}\Gamma^{\mu\nu\rho}_{3g,0}(p_1,p_2,p_3) &= ig\,f_{abc}\,Z_{3g}   \Big[ (p_1-p_2)^\rho\,\delta^{\mu\nu} + (p_2-p_3)^\mu\,\delta^{\nu\rho} + (p_3-p_1)^\nu\,\delta^{\rho\mu}\Big]\,, \label{feynman-rule-2} \\
		\Gamma^{\mu\nu\rho\sigma}_{4g,0}(p_1,p_2,p_3,p_4) &=
		- g^2  Z_{4g}\, \Big[ \,f_{abe} f_{cde} \left( \delta^{\mu\rho} \delta^{\nu\sigma} - \delta^{\nu\rho} \delta^{\mu\sigma} \right)
		+ f_{ace} f_{bde} \left( \delta^{\mu\nu} \delta^{\rho\sigma} - \delta^{\nu\rho} \delta^{\mu\sigma} \right) \nonumber \\[-1mm]
		& \qquad\qquad + f_{ade} f_{cbe} \left( \delta^{\mu\rho} \delta^{\nu\sigma} - \delta^{\mu\nu} \delta^{\rho\sigma} \right) \Big]\,. \label{feynman-rule-3}
	\end{align}

	\caption{Feynman rules for the tree-level ghost-gluon, three-gluon and four-gluon vertex}
	\label{tab:tree-level}

\end{table*}

The full four-gluon vertex depends on 136 Lorentz tensors~\cite{Eichmann:2015nra} and five color structures.
Here we keep again only the tree-level tensor
with a general dressing function $F_{4g}$ obtained by the replacement
\begin{equation}
	Z_{4g} \to F_{4g}(p_1^2,p_2^2,p_3^2,p_4^2)\,.
\end{equation}
Also in this case we assume that the dressing function only depends on the symmetric variable:
\begin{equation}\label{4gv-approx}
	F_{4g}(p_1^2,p_2^2,p_3^2,p_4^2) \approx F_{4g}\left( \frac{p_1^2+p_2^2+p_3^2+p_4^2}{4}\right).
\end{equation}

\subsection{Ghost and gluon DSEs: Single-loop terms}

In the following we work out the self-energy diagrams in Fig.~\ref{fig-dses-mom-routing-1} explicitly.
We work in Landau gauge ($\xi=0$) where the ghost-gluon vertex is UV-finite and thus we set $\widetilde Z_\Gamma=1$.

In hyperspherical variables,
the Lorentz-invariant integral measure $\int_k = \int d^4k/(2\pi)^4$ takes the form
\begin{equation}\label{int-measure}
	\begin{split}
		\int_k &= \frac{1}{(2\pi)^4}\,\frac{1}{2}\int_0^{\Lambda^2} dk^2\,k^2 \int_{-1}^1 dz\sqrt{1-z^2} \int_{-1}^1 dy \int_0^{2\pi} d\psi \\
		&= \frac{1}{(4\pi)^2}\,\int_0^{\Lambda^2} dk^2\,k^2\,\frac{2}{\pi}\int_{-1}^1 dz\sqrt{1-z^2} \,,
	\end{split}
\end{equation}
where $\Lambda$ is the hard cutoff.
Since the only Lorentz invariants appearing in the single-loop diagrams are $Q^2$, $k^2$ and $\omega = k\cdot Q = \sqrt{k^2}\sqrt{Q^2}\,z$,
the integrations over the variables $y$ and $\psi$ become trivial.
This corresponds to the frame where
\begin{equation}\label{4-momenta}
	Q = \sqrt{Q^2} \left[\begin{array}{c} 0 \\ 0 \\ 0 \\ 1 \end{array}\right], \quad
	k = \sqrt{k^2} \left[ \begin{array}{c} 0 \\ 0 \\ \sqrt{1-z^2}  \\ z \end{array}\right].
\end{equation}
The internal loop momenta in Fig.~\ref{fig-dses-mom-routing-1}
are $k^\mu_\pm = k^\mu \pm Q^\mu/2$ such that $k_\pm^2 = k^2 + Q^2/4 \pm \omega$.

\bigskip

{\tiny$\blacksquare$} The \textbf{ghost self-energy} becomes
\begin{align}
	\Sigma_G(Q^2) =& - \int_k \left[ -ig\,f_{adc}\,Q^\mu\right]\left[ -ig\,f_{cdb}\,k_+^\nu\right]  \nonumber \\
	& \times  D_G(k_+^2)\,D^{\mu\nu}(k_-)\,,
\end{align}
where the sign in front comes from the DSE (the self-energy appears with a minus sign).
With Eqs.~(\ref{ghost-app}--\ref{gluon-app}) and $f_{acd}\, f_{bcd} = N_c\,\delta_{ab}$,
the resulting expression is
\begin{equation}\label{ghost-self-energy-LG}
	\Sigma(Q^2) = -g^2 N_c \int_k \frac{G(k_+^2)}{k_+^2\,k_-^4}\, Z(k_-^2)\,k^2\,(1-z^2)\,.
\end{equation}
Note that the self-energy is negative, as it should be according to the DSE~\eqref{ghost-dse-app}: the ghost dressing function
is enhanced compared to the tree-level expression and the inverse dressing function is suppressed.
For $Q^2 = 0$, the self-energy integral becomes constant both in the scaling and decoupling case.

\bigskip

{\tiny$\blacksquare$} The \textbf{ghost loop} in the gluon DSE is given by
\begin{align}
	\Pi_\text{gh}^{\mu\nu} &= \int_k \left[ -ig\,f_{dac}\,k_-^\mu\right] \frac{G(k_+^2)}{k_+^2}\,\frac{G(k_-^2)}{k_-^2}\left[ -ig\, f_{cbd}\,k_+^\nu\right]  \nonumber \\
	&= g^2\,N_c\,\delta_{ab}\,\int_k \frac{G(k_+^2)\,G(k_-^2)}{k_+^2\,k_-^2}\,k_-^\mu\,k_+^\nu\,.
\end{align}
In principle there are two signs in front, one from the minus of the self-energy and the other from the closed fermion loop, which cancel out.
The sign of $-f_{acd}\,f_{bcd} = -N_c\,\delta_{ab}$ finally cancels the sign coming from the vertices.
By applying Eqs.~(\ref{projector-general}--\ref{projectors}) with $\lambda=0$ we extract the transverse and gauge parts:
\begin{align}
	\Pi_\text{gh}(Q^2) &= g^2 N_c\, \int_k \frac{G(k_+^2)\,G(k_-^2)}{k_+^2\,k_-^2} \left[ \frac{1}{4} + \frac{k^2 (1-4z^2)}{3Q^2}\right], \nonumber\\
	\PiT_\text{gh}(Q^2) &= g^2 N_c\, \int_k \frac{G(k_+^2)\,G(k_-^2)}{k_+^2\,k_-^2}\left[ k^2 z^2 - \frac{Q^2}{4}\right]. \label{ghost-loop}
\end{align}
In principle there are also terms $\sim z$ in the bracket, but these vanish after integration since the integral over $z$ is symmetric.
Note that the term with $(1-4z^2)$ does not produce a quadratic divergence at large $k^2$ or singularities at small $Q^2$:
in that case one has $k_\pm^2 \approx k^2$ and the integration over $z$ yields
\begin{equation}\label{1-4z2}
	\int_{-1}^1 dz \sqrt{1-z^2}\,(1-4z^2) = 0\,.
\end{equation}
Thus, only the gauge part $\PiT_\text{gh}$ has a quadratic divergence whereas the transverse part $\Pi_\text{gh}$ is only logarithmically divergent.
Its contribution to the gluon DSE is positive,
so it has the tendency to \textit{reduce} the gluon dressing function compared to its tree-level expression.

For $Q^2\to 0$ and the scaling solution,
both $\Pi_\text{gh}$ and $\PiT_\text{gh}/Q^2$ diverge with the same power $(Q^2)^{-2\kappa}$,
whereas for the decoupling solutions $\Pi_\text{gh}$ diverges logarithmically but $\PiT_\text{gh}$ remains finite.

\bigskip

{\tiny$\blacksquare$} The \textbf{gluon loop} in the gluon DSE reads
\begin{equation}
	\begin{split}
		\Pi_\text{gl}^{\mu\nu} =& -\frac{1}{2} \int_k \Big[ ig\,f_{cda}\,Z_{3g}\,\Gamma^{\rho\sigma\mu}_{3g,0}(k_+,-k_-,-Q)\Big] \\[-1mm]
		& \qquad \times D^{\rho\rho'}(k_+)\,D^{\sigma\sigma'}(k_-) \\
		& \qquad \times \left[ ig\,f_{dcb}\,\Gamma^{\sigma'\rho'\nu}_{3g}(k_-,-k_+,Q)\right],
	\end{split}
\end{equation}
where the minus in front comes from the sign of the self-energy and the factor $1/2$ from the symmetrization.
Taking again the projections and abbreviating $Z_\pm = Z(k_\pm^2)$ and $F_{3g} = F_{3g}(k_-^2,k_+^2,Q^2)$, we arrive at
\begin{equation}\label{gluon-loop}
	\begin{split}
		\Pi_\text{gl}(Q^2) &= -\frac{g^2}{2} N_c \,Z_{3g}\int_k \frac{F_{3g} \,Z_+ Z_-}{k_+^4\,k_-^4}\, K_\text{gl}\,, \\
		\PiT_\text{gl}(Q^2) &= -\frac{g^2}{2} N_c \,Z_{3g}\int_k \frac{F_{3g} \,Z_+ Z_-}{k_+^4\,k_-^4}\,\widetilde K_\text{gl}
	\end{split}
\end{equation}
with the kernels
\begin{equation}\label{gluon-loop-gauge-part}
	\begin{split}
		K_\text{gl} &= \frac{4k^6}{Q^2}\,(1-4z^2) + \frac{2}{3}\,k^4\,(16z^4-20z^2+13)\\
		& + \frac{3}{4}\,k^2\,Q^2\,(3-4z^2)\,, \\
		\widetilde K_\text{gl} &= 12 k^2 z^2\left[ k_+^2\,k_-^2 - \frac{1}{3}\,k^2\,Q^2\,(1-z^2)\right]
	\end{split}
\end{equation}
and
\begin{equation}
	k_+^2\,k_-^2 = \left(k^2 + \frac{Q^2}{4}\right)^2 - k^2\,Q^2\,z^2\,.
\end{equation}
Once again, quadratic divergences only appear in the gauge part $\PiT_\text{gl}$
whereas the transverse part $\Pi_\text{gl}$ is only logarithmically divergent due to the factor $(1-4z^2)$.
It is negative and has the tendency to \textit{increase} the gluon dressing function.
For $Q^2\to 0$, $\Pi_\text{gl}$ and $\PiT_\text{gl}$ become constant for both scaling and decoupling solutions.

\begin{figure*}
	\begin{center}
		\includegraphics[width=0.7\textwidth]{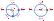}
	\end{center}
	\vspace{-5mm}
	\caption{Kinematics in the squint diagram (left) and sunset diagram (right)}\label{fig:2-loop}
\end{figure*}

\bigskip

{\tiny$\blacksquare$} The \textbf{tadpole} diagram is given by
\begin{equation}\label{tadpole}
	\Pi_\text{tad}^{\mu\nu} = -\frac{1}{2} \int_k  \Gamma^{\mu\nu\rho\sigma}_\text{4g,0}\,D^{\rho\sigma}(k)\,,
\end{equation}
which is very simple to work out since only the tree-level four-gluon vertex appears:
\begin{align}
	\Pi_\text{tad}(Q^2) &= g^2\,Z_{4g}\,N_c\,\frac{1}{3Q^2}\int_k \frac{Z(k^2)}{k^2}\,(1-4z^2) = 0\,,  \nonumber \\
	\widetilde\Pi_\text{tad}(Q^2) &= g^2\,Z_{4g}\,N_c\,\frac{9}{4} \int_k \frac{Z(k^2)}{k^2}\,. \label{tadpole}
\end{align}
The transverse part vanishes due to the factor $(1-4z^2)$, cf.~Eq.~\eqref{1-4z2}, which
is not only true in Landau gauge but also for a general gauge parameter $\xi$.
Thus, the tadpole only gives a contribution to the gauge part, which is a constant and quadratically divergent.
Because in Scenario A the gauge part is replaced by a constant, and in Scenario B
a constant is subtracted from the gauge part by Eq.~\eqref{Pi-Sc-B}, the tadpole drops out from all equations.

\subsection{Gluon DSE: Two-loop terms}

The two-loop terms in the gluon DSE are small compared to the one-loop terms (see Fig.~\ref{fig-Zinv}).
Nevertheless, they complete the DSE and ensure that the ghost and gluon propagators are two-loop exact in the UV.
In addition to the loop momentum $k$ from Eq.~\eqref{4-momenta}, we write the second loop momentum $l$ as
\begin{equation}
	l = \sqrt{l^2} \left[ \begin{array}{l} 0  \\ \sqrt{1-{z'}^2}\,\sqrt{1-y^2} \\ \sqrt{1-{z'}^2}\,y \\ z' \end{array}\right].
\end{equation}
The propagator momenta are then given by $l_\pm = l \pm k/2$ and $k' = k-Q$ as shown in Fig.~\ref{fig:2-loop}.
In this case the innermost integration is trivial, so that
there are in total five integrations instead of two. The integral measure is
\begin{equation}
	\begin{split}
		\int_k\!\!\int_l &= \frac{2}{(2\pi)^6}\,\frac{1}{4} \int dk^2\,k^2 \int dl^2 \,l^2 \\
		& \quad \times \int_{-1}^1 dz\,\sqrt{1-z^2} \int_{-1}^1 dz'\,\sqrt{1-{z'}^2} \int_{-1}^1 dy\,.
	\end{split}
\end{equation}
This would lead to a substantially higher computational demand,
but with the simplifications for the three- and four-gluon vertices
explained around Eqs.~\eqref{3gv-approx} and~\eqref{4gv-approx}
the integrands of the two-loop terms simplify substantially.
To this end, we define the variables
\begin{equation}
	\begin{split}
		\Omega(z,z',y) & = \hat{k}\cdot\hat{l} = zz'+y\sqrt{1-z^2}\sqrt{1-{z'}^2}\,, \\
		w & = \frac{z' - z\,\Omega(z,z',y)}{\sqrt{1-z^2}\,\sqrt{1-\Omega(z,z',y)^2}}\,,
	\end{split}
\end{equation}
where a hat denotes a normalized momentum,
with the inverse relations
\begin{equation}
	\begin{split}
		z'(z,\Omega,w) &= z\,\Omega + w\sqrt{1-z^2}\sqrt{1-\Omega^2}\,, \\
		y &= \displaystyle \frac{\Omega - z\,z'(z,\Omega,w)}{\sqrt{1-z^2}\sqrt{1-z'(z,\Omega,w)^2}}\,.
	\end{split}
\end{equation}
Then we can rewrite the two innermost integrations as
\begin{equation}\label{zy-womega}
	\int_{-1}^1 dz'\,\sqrt{1-{z'}^2} \int_{-1}^1 dy = \int_{-1}^1 d\Omega \,\sqrt{1-\Omega^2} \int_{-1}^1 dw\,.
\end{equation}
Below we will see that the integrands only depend on $\Omega$ but not on $w$,
so that the $w$ integration becomes trivial.
In addition, with some rearrangements of the integrals it turns out that
the squint diagram requires the same computational demand as the one-loop terms
and the sunset diagram  involves only one additional integration.

\bigskip

{\tiny$\blacksquare$}    The \textbf{squint diagram} gives the dominant contribution to the two-loop terms.
It reads
\begin{equation}
	\begin{split}
		\Pi^{\mu\nu}_\text{sq} &= -\frac{1}{2} \int_k \!\!\int_l \left[ \Gamma_{4g,0}\right]^{\mu\alpha\beta\gamma}_{ai,jk}
		\left[\Gamma_{3g}(-l_+,k,l_-)\right]^{\alpha'\delta'\beta'}_{ilj} \\
		& \quad \times               \left[\Gamma_{3g}(Q,k',-k)\right]^{\nu\gamma'\delta}_{bkl}\,
		T^{\alpha\alpha'}_{l_+}\,T^{\beta\beta'}_{l_-}\,T^{\gamma\gamma'}_{k'}\,T^{\delta\delta'}_k \\[1mm]
		& \quad \times              \frac{Z(l_+^2)}{l_+^2}\,\frac{Z(l_-^2)}{l_-^2}\,\frac{Z(k^2)}{k^2}\,\frac{Z({k'}^2)}{{k'}^2}\,
	\end{split}
\end{equation}
where the prefactor $-1/2$ comes from the minus sign in the DSE and the symmetrization,
and the transverse projectors from the gluon propagators were defined below Eq.~\eqref{gluon-app}.
We absorb all Lorentz-invariant dressing functions into a quantity $I_\text{sq}(l^2,k^2,Q^2,z,\Omega)$ defined by
\begin{equation}
	I_\text{sq} = F_{3g}(x_1)\,F_{3g}(x_2)\,\frac{Z(l_+^2)}{l_+^4}\,\frac{Z(l_-^2)}{l_-^4}\,\frac{Z(k^2)}{k^4}\,\frac{Z({k'}^2)}{{k'}^4}\,,
\end{equation}
where the three-gluon vertices depend on the symmetric variables
\begin{equation}
	\begin{split}
		x_1 &= \frac{Q^2+k^2+{k'}^2}{3} = \frac{2}{3}\,(k^2 + Q^2 - k\cdot Q)\,, \\
		x_2 &= \frac{l_+^2+l_-^2+k^2}{3} = \frac{2l^2}{3} + \frac{k^2}{2}\,.
	\end{split}
\end{equation}
The squint diagram then becomes
\begin{equation}
	\begin{split}
		\Pi^{\mu\nu}_\text{sq} &= -\frac{1}{2} \,g^4 Z_{4g}\,\frac{27}{2}\,\delta_{ab} \int_k \!\! \int_l I_\text{sq}\,K_\text{sq}^{\mu\nu} \,,
	\end{split}
\end{equation}
where the factor $27/2$ is the color trace and
the remaining Lorentz kernel $K_\text{sq}^{\mu\nu}$ is just a kinematic expression.
With the decomposition~(\ref{projector-general}--\ref{projectors}) we can split it into
its Lorentz-invariant components:
\begin{align}
	\left[ \begin{array}{c} \Pi_\text{sq}(Q^2) \\[1mm] \widetilde\Pi_\text{sq}(Q^2) \end{array} \right] &= -\frac{1}{2} \,g^4 Z_{4g}\,\frac{27}{2}
	\int_k \!\! \int_l I_\text{sq}(l^2,k^2,Q^2,z,\Omega) \nonumber \\
	& \quad \times \left[ \begin{array}{c} K_\text{sq}(Q^2,k^2,l^2,z,z',y) \\[1mm] \widetilde K_\text{sq}(Q^2,k^2,l^2,z,z',y) \end{array}\right].
\end{align}
Using Eq.~\eqref{zy-womega}, and because $I_\text{sq}$ does not depend on the variable $w$,
we can integrate out $w$ to arrive at 
\begin{align}\label{squint-2}
	\left[ \begin{array}{c} \Pi_\text{sq}(Q^2) \\[1mm] \widetilde\Pi_\text{sq}(Q^2) \end{array} \right] &= -\frac{1}{2} \,g^4 Z_{4g}\,\frac{27}{2}\,\frac{1}{(2\pi)^6}
	\int dk^2\,k^2  \nonumber \\
	& \quad \times \int dl^2\,l^2 \int_{-1}^1 dz\,\sqrt{1-z^2} \int_{-1}^1 d\Omega\,\sqrt{1-\Omega^2} \nonumber \\
	& \quad \times I_\text{sq}(l^2,k^2,Q^2,z,\Omega) \nonumber \\[1mm]
	& \quad \times \left[ \begin{array}{c} K'_\text{sq}(Q^2,k^2,l^2,z,\Omega) \\[1mm] \widetilde K'_\text{sq}(Q^2,k^2,l^2,z,\Omega) \end{array}\right].
\end{align}
The kernels $K'_\text{sq}$ and $\widetilde K'_\text{sq}$ are given by
\begin{equation*}
	\left[ \begin{array}{c} K'_\text{sq} \\[1mm] \widetilde K'_\text{sq} \end{array}\right] =
	-\frac{1}{18}\,k^2\,l^2\,(7k^2+20l^2)\,(1-\Omega^2)
	\left[ \begin{array}{c} c(Q^2,k^2,z) \\[1mm] \widetilde c(Q^2,k^2,z) \end{array}\right]
\end{equation*}
with the coefficients
\begin{equation}
	\begin{split}
		c(Q^2,k^2,z) &= k^2\left[ \frac{6k^2}{Q^2}\,(1-4z^2) + 7 - 43 z^2\right] \\
		& \quad + \frac{kz}{Q}\left[ 5k^2\,(1+8z^2) + 9Q^2\right], \\[1mm]
		\widetilde c(Q^2,k^2,z)&= 3k^2\left[ 6k^2 z^2 + Q^2\,(1+11z^2)\right] \\
		& \quad - 3kQz\left[ 5k^2\,(1+2z^2) + 3Q^2\right].
	\end{split}
\end{equation}

Even though one must integrate over $k^2$, $l^2$, $z$ and $\Omega$,
the product of the second, third and fourth line in Eq.~\eqref{squint-2} can be written as
\begin{equation}
	\left[ \begin{array}{c} A(Q^2,k^2) \\[1mm] \widetilde A(Q^2,k^2) \end{array}\right] B(k^2)
\end{equation}
with
\begin{align}
	\left[ \begin{array}{c} A(Q^2,k^2) \\[1mm] \widetilde A(Q^2,k^2) \end{array}\right] &= \frac{Z(k^2)}{k^4} \int_{-1}^1 dz\,\sqrt{1-z^2}\,F_{3g}(x_1) \nonumber \\
	& \quad \times \frac{Z({k'}^2)}{{k'}^4} \left[ \begin{array}{c} c(Q^2,k^2,z) \\[1mm] \widetilde c(Q^2,k^2,z) \end{array}\right]
\end{align}
and
\begin{align}
	B(k^2) &= -\frac{1}{18}\,k^2 \int dl^2\,l^4\,(7k^2+20l^2)\,F_{3g}(x_2) \nonumber \\
	& \quad \times \int_{-1}^1 d\Omega\,(1-\Omega^2)^{3/2}\,\frac{Z(l_+^2)}{l_+^4}\,\frac{Z(l_-^2)}{l_-^4}\,. \nonumber
\end{align}
The functions $A$ and $\widetilde A$ only require looping over three variables ($Q^2$, $k^2$ and $z$)
and the same is true for $B(k^2)$ which does not depend on $Q^2$, hence the loops go over $k^2$, $l^2$ and $\Omega$.
Therefore, calculating the squint diagram is computationally no more expensive than the one-loop diagrams,
at least within our approximations for the three-gluon vertex.
Once again, $\Pi_\text{sq}(Q^2)$ is only logarithmically divergent and
$\PiT_\text{sq}(Q^2)$ diverges quadratically.

\begin{figure*}[t]
	\includegraphics[width=1\textwidth]{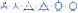}
	\caption{Momentum routing in the three-gluon vertex DSE}
	\label{fig-dses-mom-routing-2}
\end{figure*}

\bigskip

{\tiny$\blacksquare$} The \textbf{sunset diagram} is almost negligible compared to the squint diagram,
which dominates the contribution from the two-loop terms by far. It is given by
\begin{align}
	\Pi^{\mu\nu}_\text{sun} &= -\frac{1}{6} \int_k \!\!\int_l \left[ \Gamma_{4g,0}\right]^{\mu\alpha\beta\gamma}_{ai,jk}
	\left[\Gamma_{4g}(Q,l_-,k',-l_+)\right]^{\nu\gamma'\beta'\alpha'}_{bk,ji} \nonumber \\
	& \quad \times  \frac{Z(l_+^2)}{l_+^2}\,\frac{Z(l_-^2)}{l_-^2}\,\frac{Z({k'}^2)}{{k'}^2}\,
	T^{\alpha\alpha'}_{l_+}\,T^{\beta\beta'}_{k'}\,T^{\gamma\gamma'}_{l_-}\,,
\end{align}
where the prefactor $-1/6$ comes from the sign in the DSE and the symmetrization factor.
We employ the same strategy as before and absorb the dressing functions in a quantity $I_\text{sun}(l^2,k^2,Q^2,z,\Omega)$ defined by
\begin{equation}
	I_\text{sun} = F_{4g}(\overline x)\,\frac{Z(l_+^2)}{l_+^4}\,\frac{Z(l_-^2)}{l_-^4}\,\frac{Z({k'}^2)}{{k'}^4}\,,
\end{equation}
where the four-gluon vertex dressing depends on the symmetric variable
\begin{equation}
	\begin{split}
		\overline x &= \frac{l_+^2+l_-^2+{k'}^2+Q^2}{4} \\
		&= \frac{1}{2}\left( l^2 + \frac{3k^2}{4} + Q^2 - k\cdot Q\right).
	\end{split}
\end{equation}
The sunset diagram can then be written as
\begin{equation}
	\Pi^{\mu\nu}_\text{sun} = -\frac{1}{6} \,g^4 Z_{4g}\,\frac{9}{2}\,\delta_{ab} \int_k \!\! \int_l I_\text{sun}\,K_\text{sun}^{\mu\nu}\,,
\end{equation}
with the color trace $9/2$ and a kernel $K_\text{sun}^{\mu\nu}$ that is purely kinematic.
The resulting dressing functions are
\begin{align}
	\left[ \begin{array}{c} \Pi_\text{sun}(Q^2) \\[1mm] \widetilde\Pi_\text{sun}(Q^2) \end{array} \right]
	&= -\frac{1}{6} \,g^4 Z_{4g}\,\frac{9}{2}
	\int_k \!\! \int_l I_\text{sun}(l^2,k^2,Q^2,z,\Omega) \nonumber \\
	& \times \left[ \begin{array}{c} K_\text{sun}(Q^2,k^2,l^2,z,z',y) \\[1mm] \widetilde K_\text{sun}(Q^2,k^2,l^2,z,z',y) \end{array}\right].
\end{align}
Once again, $I_\text{sun}$ does not depend on the variable $w$, which can be integrated over to arrive at
\begin{align}
	\left[ \begin{array}{c} \Pi_\text{sun}(Q^2) \\[1mm] \widetilde\Pi_\text{sun}(Q^2) \end{array} \right] &= -\frac{1}{6} \,g^4 Z_{4g}\,\frac{9}{2}
	\frac{4}{(2\pi)^6}\,\frac{1}{4} \int dk^2\,k^2 \int dl^2\,l^2\, \nonumber \\
	& \times \int_{-1}^1 dz\,\sqrt{1-z^2}\,\int_{-1}^1 d\Omega \,\sqrt{1-\Omega^2} \nonumber \\
	& \times   I_\text{sun}(l^2,k^2,Q^2,z,\Omega) \nonumber \\[1mm]
	& \times \left[ \begin{array}{c} K'_\text{sun}(Q^2,k^2,l^2,z,\Omega) \\[1mm] \widetilde K'_\text{sun}(Q^2,k^2,l^2,z,\Omega) \end{array}\right]. \label{sunset-2}
\end{align}

In this case the kernels $K'_\text{sun}$ and $\widetilde K'_\text{sun}$ are more complicated, but they are still even functions
of $\Omega$ which involve at most quartic powers ($1$, $\Omega^2$ and $\Omega^4$). Thus we may expand them as follows:
\begin{equation}
	K'_\text{sun} = \sum_{n=0}^2 c_n(Q^2,k^2,l^2,z)\left[ (1-\Omega^2)\,l^2 \right]^n\,,
\end{equation}
and likewise for $\widetilde K'_\text{sun}$. The product of the second, third and fourth line in Eq.~\eqref{sunset-2} then takes the form
\begin{equation}
	\sum_{n=0}^2 l^{2n} \left[ \begin{array}{c} A_n(Q^2,k^2,l^2) \\[1mm] \widetilde A_n(Q^2,k^2,l^2) \end{array}\right]B_n(k^2,l^2)\,,
\end{equation}
where
\begin{align}
	\left[ \begin{array}{c} A_n(Q^2,k^2,l^2) \\[1mm] \widetilde A_n(Q^2,k^2,l^2) \end{array}\right] &=
	\int_{-1}^1 dz\,\sqrt{1-z^2}\,F_{4g}(\overline x)\,\frac{Z({k'}^2)}{{k'}^4} \nonumber \\
	&\quad \times \left[ \begin{array}{c} c_n(Q^2,k^2,l^2,z) \\[1mm] \widetilde c_n(Q^2,k^2,l^2,z) \end{array}\right]\,, \\
	B_n(k^2,l^2) &= \int_{-1}^1 d\Omega\,(1-\Omega^2)^{n+\frac{1}{2}}\,\frac{Z(l_+^2)}{l_+^4}\,\frac{Z(l_-^2)}{l_-^4}\,. \nonumber
\end{align}
The coefficients $c_n$ read explicitly:
\begin{align}
	c_0 &= -\frac{1}{8}\,(k^2-4l^2)^2\bigg[ \frac{6k^2}{Q^2}\,(1-4z^2)  \nonumber \\
	& \quad + \frac{4kz}{Q}\,(1+8z^2) + (1-15z^2-4z^4)\bigg]\,, \nonumber\\
	c_1 &= \frac{k^2}{6Q^2}\,(1-4z^2)\left[ -35 k^2+92 l^2\right] \\
	& \quad   - \frac{kz}{Q}\left[ \frac{k^2}{3}\,(13+92 z^2) + 4l^2\,(5-28z^2)\right] \nonumber\\
	& \quad - \frac{k^2}{6}\,(11-84 z^2 - 32 z^4) + 2l^2\,(5 - 12z^2 - 16 z^4)\,, \nonumber\\
	c_2 &= -\frac{8k^2}{3Q^2}\,(1-4z^2) + \frac{16 k z}{Q}\,(1-2z^2) \nonumber\\
	& \quad + \frac{8}{5}\,(1-12z^2+16z^4)\,. \nonumber
\end{align}
Also in this case, the potentially dangerous terms proportional to $k^2/Q^2$ do not produce quadratic divergences since they come with factors $(1-4z^2)$,
and the terms proportional to $k/Q$ are multiplied by $z$ so that their integrals vanish for $k^2\to\infty$ or $l^2\to \infty$.
The coefficients entering in $\PiT_\text{sun}(Q^2)$, on the other hand, do lead to quadratic divergences:
\begin{align}
	\widetilde c_0 &= \frac{3}{8}\,(1-z^2) (k^2-4l^2)^2 \left[ 6k^2-8kQz + Q^2(4+z^2)\right], \nonumber \\[1mm]
	\widetilde c_1 &= \frac{7}{2}\,k^4\,(8-5z^2) - 2k^2 \,l^2\,(8-23z^2) \nonumber\\
	& \quad + \frac{1}{2}\,k^2\,Q^2 (47-18z^2-8z^4) \\
	& \quad - 6l^2\,Q^2\,(1-2z^2-4z^4) \nonumber\\
	& \quad + kQz\left[ -k^2(44-23z^2) + 12 l^2 \,(2-7z^2)\right], \nonumber\\[1mm]
	\widetilde c_2 &= \frac{2}{5}\,\Big[ 5k^2\,(1-4z^2) - 30 kQz\,(1-2z^2) \nonumber\\
	& \quad - 3Q^2\,(1-12z^2+16 z^4)\Big]\,. \nonumber
\end{align}

We note that
the $\{ A_n, \widetilde A_n\}$ do not depend on $\Omega$ and
the $B_n$ do not depend on $z$ and $Q^2$.
Thus, we still saved two integrations such that
the sunset diagram only requires one additional integration compared to the one-loop and squint diagrams.

\subsection{Three-gluon vertex DSE}

Because we back-couple the three-gluon vertex into the propagator DSEs, we must also work out its own DSE.
It is given by
\begin{equation} \label{dse-3gv-0}
	\Gamma^{\mu\nu\rho}_{3g}(p_1,p_2,p_3) = \Gamma^{\mu\nu\rho}_{3g,0}(p_1,p_2,p_3) + \mM^{\mu\nu\rho}(p_1,p_2,p_3)
\end{equation}
and the corresponding diagrams are shown in Fig.~\ref{fig-dses-mom-routing-2}.
Here we neglected further two-loop diagrams as well as those containing higher $n$-point functions (such as the two-ghost-two-gluon vertex).
We also restrict ourselves to the symmetric limit, where
$p_1^2 = p_2^2 = p_3^2 = Q^2$, and to the classical tensor structure from Eq.~\eqref{feynman-rule-3}.
As shown in Ref.~\cite{Eichmann:2014xya}, this is a good approximation
since the angular dependence is mild and the effect from higher tensors are small.
A symmetrization of the diagrams is not necessary in the symmetric limit
where all six permutations are identical, hence we contract two of the three swordfish
diagrams from Fig.~\ref{fig-ym} into one (the factor 2 is implicit).

The full three-gluon vertex in the symmetric limit has the general form
\begin{equation}\label{3gv-symm}
	\Gamma^{\mu\nu\rho}(p_1,p_2,p_3) = \sum_{i=1}^3 F_{3g}^{(i)}(Q^2)\,K_i^{\mu\nu\rho}
\end{equation}
and depends on three fully antisymmetric tensors,
\begin{equation}\label{3gv-tensors}
	\begin{split}
		K_1^{\mu\nu\rho} &= q_1^\mu\,\delta^{\nu\rho} + q_2^\nu\,\delta^{\rho\mu} + q_3^\rho\,\delta^{\mu\nu}\,, \\[1mm]
		K_2^{\mu\nu\rho} &= q_1^\mu\,q_2^\nu\,q_3^\rho\,, \\[1mm]
		K_3^{\mu\nu\rho} &= q_1^\mu\,p_2^\nu\,p_3^\rho + p_1^\mu\,q_2^\nu\,p_3^\rho + p_1^\mu\,p_2^\nu\,q_3^\rho\,,
	\end{split}
\end{equation}
where $q_i = p_j - p_k$ and $\{i,j,k\}$ is an even permutation of $\{1,2,3\}$.
$K_1$ is the classical tensor, whereas the remaining ones carry higher momentum powers
and thus their dressing functions are subleading.
Moreover, $K_3$ vanishes upon transverse projection and drops out from the coupled DSEs
in Landau gauge where it is always contracted with two transverse gluons in the loops.

Eq.~\eqref{3gv-symm} can be verified as follows.
In general, the three-gluon vertex has 14 possible tensors,
which are arranged in Table V of Ref.~\cite{Eichmann:2014xya}
in terms of singlets, antisinglets and doublets under the permutation group $S_3$.
Since the full vertex is Bose-symmetric and the color structure $f_{abc}$  antisymmetric,
the Lorentz part must be antisymmetric as well. The three tensors $K_1$, $K_2$ and $K_3$ are already antisymmetric
(they correspond to $\mA'(\psi_1)$, $\mA'(\psi_2)$ and $\mA'(\psi_4)$ in the table),
whereas the remaining ones would have to be
combined with the momentum doublet $\mD=\{a,s\}$ to construct antisymmetric tensors, where
the variables $a$ and $s$ are given by
\begin{equation}\label{S0-a-s}
	a = \sqrt{3}\,\frac{p_2^2-p_1^2}{p_1^2+p_2^2+p_3^2}\,,  \quad
	s = \frac{p_1^2+p_2^2-2p_3^2}{p_1^2+p_2^2+p_3^2}\,.
\end{equation}
However, these variables vanish in the symmetric limit and thus $\mD=0$,
so that in the symmetric limit only the three tensors above survive.

In the following we work out $\mM_\text{diag}^{\mu\nu\rho}(p_1,p_2,p_3)$,
where `diag' stands for the ghost triangle, gluon triangle and the swordfish diagrams in Fig.~\ref{fig-dses-mom-routing-2}.
In practice we expand each diagram in the full basis
\begin{equation}
	\mM_\text{diag}^{\mu\nu\rho}(p_1,p_2,p_3) = \sum_{i=1}^3 \mM_\text{diag}^{(i)}(Q^2)\,K_i^{\mu\nu\rho}\,,
\end{equation}
and read off the coefficient $\mM_\text{diag}^{(1)}(Q^2)$ for the tree-level tensor $K_1^{\mu\nu\rho}$ by inverting the matrix equation
\begin{equation}\label{3gv-matrix-eq}
	\left[ K_i^{\mu\nu\rho}\,\mM_\text{diag}^{\mu\nu\rho}\right] =  \sum_{j=1}^3 \left[K_i^{\mu\nu\rho}\,K_j^{\mu\nu\rho}\right] \mM_\text{diag}^{(j)}(Q^2)\,.
\end{equation}
Note that $\mM_\text{diag}^{(1)}(Q^2)$ is different from the result obtained when taking the trace with the tree-level tensor:
\begin{equation}
	\frac{K_1^{\mu\nu\rho}\,\mM_\text{diag}^{\mu\nu\rho}}{K_1^{\mu\nu\rho}\,K_1^{\mu\nu\rho}}  =
	\mM_\text{diag}^{(1)}  -\left( 3\mM_\text{diag}^{(2)} + \mM_\text{diag}^{(3)} \right)\frac{Q^2}{6}\,,
\end{equation}
but we find that the difference is numerically negligible (which confirms that the remaining dressing functions
are strongly suppressed).
Omitting the superscript (1), the DSE~\eqref{dse-3gv-0} then becomes
\begin{equation}\label{3gv-sum}
	F_{3g}(Q^2) = Z_{3g} + \mM_\text{gh}(Q^2) + \mM_\text{gl}(Q^2) + \mM_\text{sf}(Q^2)\,.
\end{equation}

Starting from the kinematics in Fig.~\ref{fig-dses-mom-routing-2}, we define two external momenta $Q$ and $p$ and a loop momentum $k$
such that the three incoming momenta $p_i$ and the internal momenta $k_i$ are given by
\begin{equation}\renewcommand{\arraystretch}{2.2}
	\begin{array}{rl}
		p_1 &= \displaystyle -p - \frac{Q}{2}\,, \\
		p_2 &= \displaystyle p - \frac{Q}{2}\,, \\
		p_3 &= Q\,,
	\end{array}\qquad
	\begin{array}{rl}
		k_1 &= \displaystyle -k + \frac{Q}{2}\,, \\
		k_2 &= \displaystyle -k - \frac{Q}{2}\,, \\
		k_3 &= p-k\,.
	\end{array}
\end{equation}
In the symmetric limit we have $p_1^2=p_2^2=p_3^2=Q^2$ and thus $p^2 = 3Q^2/4$ and $p\cdot Q=0$,
so we can choose the four-momenta as
\begin{equation}
	\begin{split}
		Q &= \sqrt{Q^2} \left[\begin{array}{c} 0 \\ 0 \\ 0 \\ 1 \end{array}\right], \quad
		p = \frac{\sqrt{3Q^2}}{2} \left[\begin{array}{c} 0 \\ 0 \\ 1 \\ 0 \end{array}\right], \\
		k &= \sqrt{k^2} \left[ \begin{array}{l} \sqrt{1-z^2}\,\sqrt{1-y^2}\,\sin\psi \\ \sqrt{1-z^2}\,\sqrt{1-y^2}\,\cos\psi \\ \sqrt{1-z^2}\,y \\ z \end{array}\right].
	\end{split}
\end{equation}
The arguments of the internal propagators then become
\begin{equation}\label{3gv-momenta}\renewcommand{\arraystretch}{2.2}
	\begin{split}
		k_1^2 &= k^2 + \frac{Q^2}{4} - kQ\,z\,, \\
		k_2^2 &= k^2 + \frac{Q^2}{4} + kQ\,z\,, \\
		k_3^2 &= k^2 + \frac{3Q^2}{4} - \sqrt{3}kQ\,y\sqrt{1-z^2}\,,
	\end{split}
\end{equation}
where we wrote $Q=\sqrt{Q^2}$ and $k=\sqrt{k^2}$.
The  symmetric variables that enter as arguments of the internal three-gluon vertices are given by
\begin{equation}\label{3gv-momenta}\renewcommand{\arraystretch}{2.2}
	\begin{split}
		x_1 &= \frac{Q^2 + k_2^2 + k_3^2}{3} \,, \\
		x_2 &= \frac{k_1^2 + Q^2 + k_3^2}{3}\,, \\
		x_3 &= \frac{k_1^2 + k_2^2 + Q^2}{3} = \frac{4k^2+3Q^2}{6}\,,
	\end{split}
\end{equation}
and the symmetric variable that enters in the four-gluon vertex is
\begin{equation}
	x_4 = \frac{2Q^2 + k_1^2 + k_2^2}{4} = \frac{4k^2+5Q^2}{8}\,.
\end{equation}
The integral measure has the same form as in Eq.~\eqref{int-measure} except that the integration over $y$ is no longer trivial:
\begin{equation}\label{int-measure-2}
	\int_k = \frac{1}{(2\pi)^3}\,\frac{1}{2}\int_0^{\Lambda^2} dk^2\,k^2 \int_{-1}^1 dz\sqrt{1-z^2} \int_{-1}^1 dy \,.
\end{equation}
Note that $k_1^2$, $k_2^2$ do not depend on $y$  and $x_3$, $x_4$ depend neither on $z$ nor $y$.
As before, we restrict ourselves to Landau gauge with $\xi=0$ and $\widetilde Z_\Gamma = 1$.

\bigskip

{\tiny$\blacksquare$}   The \textbf{ghost triangle} is given by
\begin{equation}
	\begin{split}
		&ig f_{abc} \mM_\text{gh}^{\mu\nu\rho} = -2 \int_k \left[ -ig f_{b'ac'}\,k_2^\mu\right]\left[ -ig f_{c'ba'}\,k_3^\nu\right]\\
		& \qquad \times \left[ -ig f_{a'cb'}\,k_1^\rho\right]
		D_G(k_1^2)\,D_G(k_2^2)\,D_G(k_3^2)\,,
	\end{split}
\end{equation}
where the symmetry factor 2 comes from the Lagrangian and the minus from the closed fermion loop.
The color factors combine to $-(N_c/2)\,f_{abc}$,
and with a minus from each ghost propagator we arrive at
\begin{equation}
	\mM_\text{gh}^{\mu\nu\rho} = -g^2N_c \int_k \frac{G(k_1^2)\,G(k_2^2)\,G(k_3^2)}{k_1^2\,k_2^2\,k_3^2} \,k_2^\mu\,k_3^\nu\,k_1^\rho\,.
\end{equation}
With the projection~\eqref{3gv-matrix-eq},
the component of the tree-level tensor becomes
\begin{equation}\label{ghost-triangle}
	\begin{split}
		\mM_\text{gh}(Q^2) &= -g^2 N_c \int_k \frac{G(k_1^2)\,G(k_2^2)\,G(k_3^2)}{k_1^2\,k_2^2\,k_3^2} \\
		& \quad \times \frac{k^2}{12}\,(1-z^2)(1-y^2)\,.
	\end{split}
\end{equation}
The ghost triangle is negative since the integrand is always positive.
In practice the ghost contribution is small, except at IR momenta where it
behaves like $\mM_\text{gh}(0) \sim -\int dk^2\,G(k^2)^3/k^2$.
For the decoupling case this produces a logarithmic divergence, whereas for the scaling case the diagram scales with the power $(Q^2)^{-3\kappa}$.
In both cases the ghost triangle dominates the infrared (see e.g. Fig.~8 in~\cite{Eichmann:2014xya}), and because the tree-level term is positive,
the three-gluon vertex necessarily has a zero crossing at intermediate momenta.

We also note that if we had implemented the full ghost-gluon vertex from Eq.~\eqref{ggl-vertex},
then by the projection~\eqref{3gv-matrix-eq} the longitudinal $B$ term
(which may have massless poles)  contributes only to $\mM_\text{gh}^{(3)}(Q^2)$,
which is the dressing function of the tensor $K_3$ in Eq.~\eqref{3gv-tensors}.
However, $K_3$ is longitudinal and drops out from the DSEs in Landau gauge because
it is contracted with two internal transverse gluon lines in each loop diagram where the three-gluon vertex appears.
Therefore, the longitudinal poles from the ghost-gluon vertex do not couple
into the three-gluon vertex, at least not in the symmetric limit. For the same reason
$K_3$ would drop out from the gluon DSE and the ghost-gluon vertex DSE in Fig.~\ref{fig-ggl-dse}.
In our setup, longitudinal poles can thus only emerge from the ghost-gluon vertex.

\bigskip

{\tiny$\blacksquare$}     For the \textbf{gluon triangle} the color factor is the same apart from a minus sign, cf.~Eq.~\eqref{feynman-rule-2},
and there is no prefactor in front of the integral. This yields
\begin{equation}
	\begin{split}
		\mM_\text{gl}^{\mu\nu\rho} &= g^2\frac{N_c}{2} \int_k \frac{Z(k_1^2)\,Z(k_2^2)\,Z(k_3^2)}{k_1^2\,k_2^2\,k_3^2} \\
		& \quad \times \Gamma_{3g,0}^{\overline{\beta\alpha}\rho}(k_2,-k_1,p_3)
		\,\Gamma_{3g}^{\overline{\alpha\gamma}\nu}(k_1,-k_3,p_2) \\
		& \quad \times \Gamma_{3g}^{\overline{\gamma\beta}\mu}(k_3,-k_2,p_1)\,,
	\end{split}
\end{equation}
where the indices under the bar are contracted with the transverse projectors from the gluon propagators
with respect to the first two momentum arguments of each vertex.
In this case the tree-level component becomes
\begin{equation}\label{gluon-triangle}
	\begin{split}
		\mM_\text{gl}(Q^2) &= g^2\frac{N_c}{2} \,Z_{3g}\int_k \frac{Z(k_1^2)\,Z(k_2^2)\,Z(k_3^2)}{k_1^4\,k_2^4\,k_3^4} \\
		& \quad \times F_{3g}(x_1)\,F_{3g}(x_2)\,K_{gl}\,,
	\end{split}
\end{equation}
where the kernel $K_\text{gl}$ is given by
\begin{equation}
	K_\text{gl} = \sum_{n=0}^3 2k^6\left[\frac{Q^2}{k^2}\right]^n \left[ c_n \gamma k^2 + d_n y \sqrt{\frac{\gamma k^2 Q^2 }{3}} \right].
\end{equation}
Here we abbreviated $\gamma=1-z^2$ and the coefficients are
\begin{equation}\renewcommand{\arraystretch}{1.5}
	\begin{split}
		c_0 &= 1-y^2\,, \\
		c_1 &= 3-5y^2 +\frac{1}{12}\,\gamma\,(5+50y^2+9y^4) \\
		c_2 &= \frac{1}{48}\left( 87 - 39y^2 + \gamma\,(19+118y^2-9y^4) \right), \\
		c_3 &= \frac{1}{32} (11+17y^2),  \\[2mm]
		d_0 &= 6 + \gamma\,(y^2-9)\,, \\
		d_1 &= 3+\frac{1}{4}\,\gamma\,(y^2-33)-\frac{8}{3}\,\gamma^2, \\
		d_2 &= \frac{1}{8} \left( 3-\gamma\,(y^2+31)\right), \\
		d_3 &= -\frac{3}{8}\,.
	\end{split}
\end{equation}
The gluon triangle contribution to the three-gluon vertex
is usually positive and small.

\bigskip

{\tiny$\blacksquare$}
The two \textbf{swordfish diagrams} in Fig.~\ref{fig-dses-mom-routing-2} turn out to be the leading loop contributions to the three-gluon vertex DSE,
apart from the ghost triangle which dominates the IR but is otherwise small
(see Fig.~8 in~\cite{Eichmann:2014xya} and Fig.~24 in Ref.~\cite{Huber:2018ned} for similar results;
in our Setups 2 and 3 we find the gluon triangle to be even more suppressed).
Each of them has a symmetry factor $\frac{1}{2}$ and
the first diagram picks up a factor 2 because it counts twice in the symmetrization.
Using $f_{acd}\,f_{bcd} = N_c\,\delta_{ab}$ and
\begin{equation}
	f_{ab'c'}\,f_{a'bc'}\,f_{a'b'c} = \frac{N_c}{2}\,f_{abc}\,,
\end{equation}
then the combination of color factors from the
three- and four-gluon vertex results in the four-gluon vertex combination $\Gamma_1 + \Gamma_2/2$
from Eq.~\eqref{4gv-tree-doublet},
which is equivalent to an effective vertex
\begin{equation}
	\Gamma_\text{4g}^{\mu\nu\rho\sigma}(p_1,p_2,p_3,p_4) = F_\text{4g} \left(\delta^{\mu\rho}\,\delta^{\nu\sigma} - \delta^{\nu\rho}\,\delta^{\mu\sigma}\right)
\end{equation}
to be multiplied with $\frac{3}{2} g^2 N_c$.
The first swordfish diagram contains the tree-level four-gluon vertex ($F_{4g} \to Z_4$)
whereas in the second diagram the full vertex $F_{4g}(x_4)$ depends on the symmetric variable $x_4$.
Together with all prefactors, we arrive at
\begin{equation}
	\begin{split}
		\mM_\text{sf}^{\mu\nu\rho} &= g^2\,\frac{3N_c}{4}\int_k \frac{Z(k_1^2)\,Z(k_2^2)}{k_1^2\,k_2^2} \\
		\quad \times  \Big[ 2\,&\Gamma_{3g}^{\overline{\beta\alpha}\rho}(k_2,-k_1,p_3)\,\Gamma_\text{4g,0}^{\mu\nu\overline{\alpha\beta}}(p_1,p_2,k_1,-k_2)  \\
		\quad + &\Gamma_\text{3g,0}^{\overline{\beta\alpha}\rho}(k_2,-k_1,p_3)\,\Gamma_\text{4g}^{\mu\nu\overline{\alpha\beta}}(p_1,p_2,k_1,-k_2)  \Big]\,,
	\end{split}
\end{equation}
where the barred indices denote again transverse projection.
Here the tree-level component becomes
\begin{equation}\label{swordfish-LG}
	\begin{split}
		\mM_\text{sf}(Q^2) &= -g^2\frac{3N_c}{4} \int_k \frac{Z(k_1^2)\,Z(k_2^2)}{k_1^4\,k_2^4} \\
		& \quad \times \left[ 2Z_{4g}\,F_{3g}(x_3) + Z_{3g}\,F_{4g}(x_4)  \right]\,K_\text{sf}\,,
	\end{split}
\end{equation}
where the kernel is given by
\begin{equation}
	K_\text{sf} = \frac{k^2}{12}\,(1-z^2)\left[ 4k^2\,(3+y^2) + Q^2\,(5-y^2) \right].
\end{equation}
Observe that the $y$ dependence inside the integrand is only carried by $K_\text{sf}$,
so we can integrate it out and use instead
\begin{equation}
	K'_\text{sf} = \frac{1}{2}\int dy \,K_\text{sf} = \frac{k^2}{18}\,(20k^2+7Q^2)(1-z^2)\,.
\end{equation}
$K'_\text{sf}$ is positive and thus the contribution from the swordfish diagrams is usually negative.

\section{Renormalization}  \label{app:renormalization}

Here we provide details on the arguments made in Sec.~\ref{sec:renormalization-main}.
Using the expressions for the self-energies from Appendix~\ref{app:expl},
the DSEs for the ghost and gluon propagator and three-gluon vertex read
\begin{equation} \label{dses-3}
	\begin{split}
		G(Q^2)^{-1} &= Z_c + \Sigma(Q^2)\,, \\
		Z(Q^2)^{-1} &= Z_A + \mathbf{\Pi}(Q^2)\,,  \\
		F_{3g}(Q^2) &= Z_{3g} + \mM(Q^2) \,,
	\end{split}
\end{equation}
where $\mathbf{\Pi}(Q^2) = \Pi(Q^2) + \PiT(Q^2)/Q^2$.
As explained in the main text, in practice it is
convenient to renormalize the gluon DSE at an arbitrary renormalization scale $Q^2 = \mu^2$
and set $Z(\mu^2)$ as a renormalization condition,
and the ghost dressing function $G(0)$ at the origin $Q^2=0$:
\begin{equation}\label{dses-2}
	\begin{split}
		G(Q^2)^{-1} &= G_0^{-1} + \Sigma(Q^2) - \Sigma(0)\,, \\
		Z(Q^2)^{-1} &= Z_\mu^{-1} + \mathbf{\Pi}(Q^2) - \mathbf{\Pi}(\mu^2)\,, \\
		F_{3g}(Q^2) &= Z_{3g} + \mM(Q^2)\,.
	\end{split}
\end{equation}
The ghost and gluon renormalization constants are then dynamically determined from
\begin{equation}\label{rcs-2}
	Z_c = G_0^{-1} - \Sigma(0)\,, \quad
	Z_A = Z_\mu^{-1} - \mathbf{\Pi}(\mu^2)\,,
\end{equation}
and the renormalization constants $Z_{3g}$ and $Z_{4g}$ are fixed from
Eq.~\eqref{ren-constants-2}.

Suppressing all momentum dependencies and kinematic functions, the structural form of the self-energy terms~\eqref{ghost-self-energy-LG}
in the ghost DSE,
\eqref{ghost-loop}, \eqref{gluon-loop}, \eqref{tadpole}, \eqref{squint-2}, \eqref{sunset-2} in the gluon DSE and
\eqref{ghost-triangle}, \eqref{gluon-triangle}, \eqref{swordfish-LG} in the three-gluon vertex DSE is
\begin{equation}\label{check}
	\begin{split}
		\Sigma &\sim  \frac{g^2}{4\pi} \int G Z \,, \\
		\mathbf\Pi_\text{1-loop} &\sim  \frac{g^2}{4\pi} \left[\begin{array}{c} G^2 \\ Z_{3g}\,F_{3g}\,Z^2 \\ Z_{4g}\,Z\end{array}\right] , \\
		\mathbf\Pi_\text{2-loop} &\sim  \left(\frac{g^2}{4\pi}\right)^2 \int \left[\begin{array}{c} Z_{4g}\,F_{4g}\,Z^3 \\[1mm] Z_{4g}\,F_{3g}^2\,Z^4\end{array}\right] , \\
		\mM &\sim \frac{g^2}{4\pi}\int \left[\begin{array}{c} G^3 \\ Z_{3g}\,F_{3g}^2\,Z^3 \\ Z_{3g}\,F_{4g}\,Z^2 \\ Z_{4g}\,F_{3g}\,Z^2 \end{array}\right].
	\end{split}
\end{equation}
The terms under each integral must renormalize in the same way. Indeed,
from Eqs.~(\ref{ren-consts}--\ref{ren-constants-2}) we have e.g.
\begin{equation}
	\begin{split}
		g^2\,G^2 &= Z_A\left[ g^2\,G^2\right]^{(B)}\,, \\
		g^2\,Z_{3g}\,F_{3g}\,Z^2 &= Z_A \left[ g^2\,F_{3g}\,Z^2\right]^{(B)}\,, \\
		g^2\,Z_{4g}\,Z &= Z_A \left[ g^2\,Z\right]^{(B)}\,.
	\end{split}
\end{equation}

Now let us redefine the renormalized dressing functions $G$ and $Z$ by
\begin{equation}
	G' = \sqrt{\gamma}\,\frac{G}{G_0}\,, \quad
	Z' = \lambda\,\frac{Z}{Z_\mu}\,,
\end{equation}
with arbitrary factors $\gamma$ and $\lambda$,
and make the same redefinitions for all quantities that renormalize like $G$ and~$Z$:
\begin{equation*}\renewcommand{\arraystretch}{2.2}
	\begin{split}
		\left\{Z_c', \Sigma'\right\} &=  \left\{Z_c, \Sigma\right\} \frac{G_0}{\sqrt{\gamma}}\,, \\
		\left\{Z_A', \mathbf{\Pi}' \right\} &= \left\{Z_A, \mathbf{\Pi} \right\} \frac{Z_\mu}{\lambda}\,,   \\
		\left\{ F_{3g}',\,Z_{3g}', \mM'\right\} &=  \left\{ F_{3g},\,Z_{3g}, \mM \right\}\frac{\sqrt{\gamma}}{G_0}\frac{Z_\mu}{\lambda}\,, \\
		\left\{ F_{4g}',\,Z_{4g}'\right\} &=  \left\{ F_{4g},\,Z_{4g}\right\}\,\frac{\gamma}{G_0^2}\frac{Z_\mu}{\lambda}\,.
	\end{split}
\end{equation*}
Then Eq.~\eqref{check} becomes
\begin{equation}\label{check-2}
	\begin{split}
		\Sigma' &\sim  \frac{\alpha}{\gamma\lambda} \int G' Z' \,, \\
		\mathbf\Pi'_\text{1-loop} &\sim  \frac{\alpha}{\gamma\lambda}\int \left[\begin{array}{c} {G'}^2 \\ Z_{3g}'\,F_{3g}'\,{Z'}^2 \\ Z_{4g}'\,Z'\end{array}\right] , \\
		\mathbf\Pi'_\text{2-loop} &\sim  \left(\frac{\alpha}{\gamma\lambda}\right)^2\int \left[\begin{array}{c} Z_{4g}'\,F_{4g}'\,{Z'}^3 \\[1mm] Z_{4g}'\,F_{3g}^{'2}\,{Z'}^4\end{array}\right] , \\
		\mM' &\sim \frac{\alpha}{\gamma\lambda}\int \left[\begin{array}{c} {G'}^3 \\ Z_{3g}'\,F_{3g}^{'2}\,{Z'}^3 \\ Z_{3g}'\,F_{4g}'\,{Z'}^2 \\ Z_{4g}'\,F_{3g}'\,{Z'}^2 \end{array}\right],
	\end{split}
\end{equation}
where $\alpha$ is defined by
\begin{equation}\label{def-alpha-2}
	\alpha = \frac{g^2}{4\pi}\,Z_\mu\,G_0^2\,.
\end{equation}
The resulting DSEs assume the form
\begin{equation} \label{dses-4}
	\begin{split}
		G'(Q^2)^{-1} &= \frac{1}{\sqrt{\gamma}} + \Sigma'(Q^2) - \Sigma'(0)\,, \\
		Z'(Q^2)^{-1} &= \frac{1}{\lambda} + \mathbf{\Pi}'(Q^2) - \mathbf{\Pi}'(\mu^2)\,,  \\[1mm]
		F'_{3g}(Q^2) &= Z_{3g}' + \mM'(Q^2) \,, \\
		Z_c' &= \frac{1}{\sqrt{\gamma}} - \Sigma'(0)\,, \\
		Z_A' &= \frac{1}{\lambda} - \mathbf{\Pi}'(\mu^2)\,,
	\end{split}
\end{equation}
where $G_0$ and $Z_\mu$ no longer appear but instead $\alpha$, $\gamma$ and $\lambda$.
Removing the primes again and setting $\gamma=\lambda=1$, we arrive at Eq.~\eqref{dses-redef-1}
(in that case we pulled out $\alpha$ from the self-energies), whereas setting $\gamma=\alpha$ and $\lambda=1$
leads to Eq.~\eqref{dses-redef-2} (where we also redefined the scale).

For the same reason we included the factor
\begin{equation}
	\frac{g^2}{4\pi}\,G_0^2 = \frac{\alpha}{Z_\mu}
\end{equation}
in Eq.~\eqref{PiT-const}
such that $\PiT'(Q^2) = \beta\mu^2$,
because otherwise this term would not renormalize correctly.

\section{Numerics} \label{sec:numerics}

Solving the DSEs in Fig.~\ref{fig-ym} is numerically quite non-trivial
because the dressing functions can diverge both in the IR and UV,
change signs, and they generally emerge from competing contributions
which can vary substantially over many orders of magnitude.

\begin{figure}[t]
	\includegraphics[width=0.93\columnwidth]{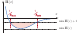}
	\caption{Sketch of the gluon self-energy and the limits on
		the subtraction point $x_\text{sub} = 1/\beta$ it imposes.}
	\label{fig-beta-limits}
\end{figure}

The prime example is the gluon DSE in Eq.~\eqref{dses-redef-2}, where
one must ensure that its r.h.s. is positive for all $x$ in each iteration step,
since a zero crossing would lead to a spacelike pole in the gluon dressing function $Z(x)$
and thus a tachyonic pole in the gluon propagator, i.e.,
\begin{equation}
	1 + \mathbf{\Pi}(x) - \mathbf{\Pi}(\tfrac{1}{\beta}) > 0 \quad \forall\,x\,.
\end{equation}
The self-energy $\mathbf{\Pi}(x)$ is sketched in Fig.~\ref{fig-beta-limits}.
It diverges in the IR; with increasing $x$ it
eventually becomes negative and develops a minimum
before at large $x$ it approaches zero from below. This leads to the condition
\begin{equation}
	\mathbf{\Pi}(\tfrac{1}{\beta}) <  \min \mathbf{\Pi}(x) + 1
\end{equation}
which defines an interval for the subtraction point $x_\text{sub} = 1/\beta$.
Thus, $\beta$ can neither become too small nor too large.
In addition, the interval can differ in the iteration steps since  $\mathbf{\Pi}(x)$
can take different intermediate values before convergence is reached.
Ideally the starting guess for $\mathbf{\Pi}(x)$
should not be too different from the converged solution to prevent the function from doing
wild jumps during the iteration.
In practice we always
scan the full $\alpha$ range from $\alpha\to 0$ up to $\alpha\to\infty$,
where we extrapolate the converged solutions at the previous values of $\alpha$
as the starting guess for the next~$\alpha$.
For small $\alpha$ convergence is easy to achieve,
whereas with increasing $\alpha$ the bump of the gluon dressing which defines
the shape in Fig.~\ref{fig-beta-limits} becomes increasingly narrow
and this limits the window in $\beta$.
For intermediate values of $\alpha$ the typical intervals  are $10^{-6} \lesssim \beta \lesssim 0.5$.

Because convergence can be very slow, we employ a Newton method whose implementation in the Yang-Mills system
was first described in Ref.~\cite{Atkinson:1997tu}.
To this end, we first map the momentum interval $x \in (0,\infty)$ to an interval $\omega \in (-1,1)$
in such a way that we can adjust the IR and UV cutoffs $\omega_\text{min}$ and $\omega_\text{max}$
and adjust the density of grid points in the IR,  UV and  mid-momentum regions separately.
Next, we expand the dressing functions $G(x)$, $Z(x)$ and $F_{3g}(x)$
in polynomials,
\begin{equation}\label{GZF-Newton}
	\begin{split}
		\ln G(x)^{-1} &= \sum_n g_n\,P_n(\omega) \,, \\
		\ln Z(x)^{-1} &= \sum_n z_n\,P_n(\omega) \,, \\
		\text{asinh}\,F_{3g}(x) &= \sum_n f_n\,P_n(\omega)\,,
	\end{split}
\end{equation}
where the logarithm and asinh ensure
that the functions do not vary overly strongly over the momentum range
and thus a polynomial expansion converges rapidly ($n_\text{max}=64$ is usually  sufficient).
In practice we employ Legendre polynomials for $P_n(\omega)$ although
the choice does not matter.
The inverse relations are
\begin{equation}
	\begin{split}
		g_n &= \int d\omega\,\Omega(\omega)\,P_n(\omega)\ln G(x)^{-1} , \\
		z_n &= \int d\omega\,\Omega(\omega)\,P_n(\omega)\ln Z(x)^{-1} , \\
		f_n &= \int d\omega\,\Omega(\omega)\,P_n(\omega) \,\text{asinh}\, F_{3g}(x)\,,
	\end{split}
\end{equation}
where $\Omega(\omega)$ are the polynomial weights according to the orthogonality relation
\begin{equation}
	\int d\omega\,\Omega(\omega)\,P_m(\omega)\,P_n(\omega) = \delta_{mn}\,.
\end{equation}

The DSEs in Eq.~\eqref{dses-redef-2} then take the form $\vect{x} = \vect{r}(\vect{x})$,
where the vector $\vect{x}=\{ g_n, z_n, f_n, Z_c, Z_A\}$ contains the moments of the dressing functions
together with the renormalization constants,
and  $\vect{r}(\vect{x})$ denotes the right-hand side of the DSEs.
The `natural iteration method' means that if $\vect{x_s}$ is the result of the current iteration,
one calculates $\vect{x}_\text{new} = \vect{r}(\vect{x_s})$,
implements the vector $\vect{x}_\text{new}$ as the new start guess in the r.h.s. and repeats, hoping that the system eventually converges.
The convergence criterion is to minimize the function
\begin{equation}\label{newton-cond}
	\vect{F}(\vect{x}) = \vect{x} - \vect{r}(\vect{x}) \stackrel{!}{=} 0  \qquad \text{or} \qquad \vect{F}(\vect{x})^2 \stackrel{!}{=} 0  \,.
\end{equation}

The Newton method tries to find a better guess for the minimum from the linear approximation
\begin{equation}
	F_i(\vect{x_s} + \vect{ \delta x}) = F_i(\vect{x_s}) + \underbrace{\left.\frac{\p F_i}{\p x_j}\right|_{\vect{x}=\vect{x_s}}}_{J_{ij}} \delta x_j + \dots \stackrel{!}{=} 0\,,
\end{equation}
where $\vect{\delta x}$ and $\vect{x}_\text{new}$ are determined from
\begin{equation}
	J\, \vect{\delta x} \stackrel{!}{=} -\vect{F}(\vect{x_s}) \quad \Rightarrow \quad \vect{x}_\text{new} = \vect{x_s} + \vect{\delta x}\,.
\end{equation}
This  finds the direction $\vect{\delta x}$ of the largest variation, following a tangent along the minimization function until it crosses zero.
We optimize the procedure using backtracking, where we test different solutions
$\vect{x}_\text{new}(\lambda) = \vect{x_s} + \lambda\,\vect{\delta x}$;
we then employ the value of $\lambda$ that minimizes $\vect{F}(\vect{x})^2$ along that direction as the starting guess for the next iteration
and repeat until the minimum is found.

In practice we still solve the DSEs for the dressing functions $G(x)$, $Z(x)$ and $F_{3g}(x)$
and only insert the Newton step between iterations: After one iteration step
in Eq.~\eqref{dses-redef-2}, we convert the solutions to the vector $\vect{F}(\vect{x_s})$ and the Jacobian matrix $J$,
determine $\vect{x}_\text{new}(\lambda)$ and convert it back to $G(x)$, $Z(x)$, $F_{3g}(x)$, $Z_c$ and $Z_A$
to be used in the next iteration. Each $\lambda$ amounts to another iteration, however with the same $J$.
Therefore, the main additional complication is the computation of $J$.

As an example, consider the ghost DSE in~\eqref{dses-redef-2}
whose self-energy is given in Eq.~\eqref{ghost-self-energy-LG}.
After redefinitions and rescaling, it becomes
\begin{equation*}
	\begin{split}
		\Sigma(x) = -\mN \int  \mathbf{K}_\Sigma\,, \quad
		\mathbf{K}_\Sigma = \frac{G(u_+)\,Z(u_-)}{u_+\,u_-^2}\, u\,(1-z^2)\,,
	\end{split}
\end{equation*}
where $\int$ denotes the (dimensionless) four-momentum integration, $\mN = 4\pi N_c$,
and $u_\pm = u + x/4 \pm \sqrt{u x}\,z$.
The ghost part of the vector $\vect{F}(\vect{x})$ in Eq.~\eqref{newton-cond} is
\begin{equation}
	\begin{split}
		F_i^\text{(gh)}
		= g_i - \int d\omega\,\Omega(\omega)\,P_n(\omega)\ln \bar G(x)^{-1}\, ,
	\end{split}
\end{equation}
where the bar denotes the outcome on the r.h.s. of the DSE.
The corresponding entries in the Jacobian are
\begin{align}
	\frac{\p F_i^\text{(gh)}}{\p g_j} &= \delta_{ij} - \int d\omega\,\Omega(\omega)\,P_i(\omega) \,\bar G(x)\,\frac{\p}{\p g_j}\left[ \Sigma(x) - \Sigma(0)\right], \nonumber\\
	\frac{\p F_i^\text{(gh)}}{\p z_j} &= \quad \;  - \int d\omega\,\Omega(\omega)\,P_i(\omega) \,\bar G(x)\,\frac{\p}{\p z_j}\left[ \Sigma(x) - \Sigma(0)\right], \nonumber\\
    \frac{\p F_i^\text{(gh)}}{\p Z_c} &= \quad \; - \int d\omega\,\Omega(\omega)\,P_i(\omega) \,\bar G(x)\,,
\end{align}
together with
\begin{equation}
   	\frac{\p F_i^\text{(gh)}}{\p f_j} = 0\,, \qquad
	\frac{\p F_i^\text{(gh)}}{\p Z_A} = 0\,.
\end{equation}
The ingredients are the derivatives of the ghost self-energy,
where the derivative with respect to $f_j$ vanishes because $\Sigma(x)$ does not depend on the three-gluon vertex.
From $\bar G(x)^{-1} = Z_c + \Sigma(x)$ we have $\p/\p Z_c\,\bar G(x)^{-1} = 1$, and the derivative with respect to $Z_A$ vanishes
because $Z_A$ does not explicitly appear in the ghost DSE.

From Eq.~\eqref{GZF-Newton} we have
\begin{equation}\label{newton-der-2}
	\begin{split}
		\frac{\p G(x)}{\p g_j} &= -P_j(\omega)\,G(x)\,, \\
		\frac{\p Z(x)}{\p z_j} &= -P_j(\omega)\,Z(x)
	\end{split}
\end{equation}
and therefore
\begin{equation}\label{ghost-self-energy-newton}
	\begin{split}
		\frac{\p}{\p g_j}\,\Sigma(x) &= \mN \int  \mathbf{K}_\Sigma\,P_j(\omega_+)\,, \\
		\frac{\p}{\p z_j}\,\Sigma(x) &= \mN \int  \mathbf{K}_\Sigma\,P_j(\omega_-)\,,
	\end{split}
\end{equation}
where the mapping $x \leftrightarrow \omega$ implies the same mapping for $u_\pm \leftrightarrow \omega_\pm$.
Thus, the derivatives of the self-energies are structurally similar to the self-energies
except for the appearance of the polynomials.
The remaining entries of the matrix $J$ are constructed
along the same lines.

Employing these techniques, the solution of the coupled DSEs for a given parameter set $(\alpha,\beta)$ typically
takes $5 \dots 10$ minutes on a single CPU,
with $10 \dots  20$ iterations until convergence is reached.

\section{Longitudinal singularities in the ghost-gluon vertex}\label{sec:long-sing}

Here we provide details on the Bethe-Salpeter equation (BSE) for the longitudinal part of the ghost-gluon vertex
discussed in Sec.~\ref{sec-long-sing-main}.
To this end, we write the general ghost-gluon vertex in Eq.~\eqref{ggl-vertex} as
\begin{equation}\label{gh-gl-vertex-2}
	\Gamma^\mu_\text{gh}(q,Q) = -ig  f_{abc}\left[ a\,q^\mu + b\,Q^\mu\right],
\end{equation}
where $q = p-Q/2$ is the average ghost momentum
and the two dressing functions $a(q^2,q\cdot Q,Q^2)$, $b(q^2,q\cdot Q,Q^2)$ are related to $A$ and $B$ by
\begin{equation}
	A = a-1\,, \quad
	B = b - \frac{a}{2}\,.
\end{equation}
From Eq.~\eqref{gh-gl-vertex-2} one can project out $a$ and $b$ via
\begin{equation}\label{gh-gl-projection}
	\begin{split}
		a &= \frac{Q^2\,q\cdot \Gamma - q\cdot Q\,Q\cdot \Gamma}{q^2\,Q^2 - (q\cdot Q)^2}\,, \\
		b &= \frac{q^2\,Q\cdot \Gamma - q\cdot Q\,q\cdot \Gamma}{q^2\,Q^2 - (q\cdot Q)^2}\,.
	\end{split}
\end{equation}
The longitudinal massless poles we are looking for can only come from the function $b$.
Under a ghost-antighost symmetry the vertex would be charge-conjugation symmetric,
$\Gamma^\mu_\text{gh}(q,Q) = -\Gamma^\mu_\text{gh}(-q,Q)$, such that $b$
would need to be antisymmetric in $q\cdot Q$, i.e., $b = (q\cdot Q)\,b'$
with $b'$ symmetric. In our present formulation of Landau gauge this symmetry
does not hold and therefore $b$ can also have even terms in $q\cdot Q$.

Writing the vertex DSE as
\begin{equation}
	a\,q^\mu + b\,Q^\mu = q^\mu + \frac{Q^\mu}{2} + \mM_\text{tot}^\mu\,,
\end{equation}
where $\mM_\text{tot}^\mu$ is the sum of all loop diagrams,
then the contribution from the `Abelian' diagram (second diagram on the r.h.s. in Fig.~\ref{fig-ggl-dse}) is given by
\begin{equation}
	\begin{split}
		\mM^\mu &= g^2 \frac{N_c}{2}\int_k (a_1\,k^\mu + b_1\,Q^\mu)\,\mI\,, \\
		\mI &= a_2\, \frac{G(k_+^2)\,G(k_-^2)}{k_+^2\,k_-^2}\,\frac{Z(l^2)}{l^4}\,K_0\,.
	\end{split}
\end{equation}
Here, $k_\pm = k\pm Q/2$ are the internal ghost momenta, $l = k-q$ is the gluon momentum,
$a_1=a(k^2,k\cdot Q,Q^2)$ and $b_1=b(k^2,k\cdot Q,Q^2)$ are the dressing functions for the top vertex, and $a_2$ is attached to the vertex on the right.
Because the internal gluons are transverse, the contribution from $b_2$ drops out.
Writing the momenta as
\begin{equation}
	\begin{split}
		Q &= \sqrt{Q^2} \left[\begin{array}{c} 0 \\ 0 \\ 0 \\ 1 \end{array}\right], \quad
		q = \sqrt{q^2} \left[\begin{array}{c} 0 \\ 0 \\ \sqrt{1-z^2} \\ z \end{array}\right], \\
		k &= \sqrt{k^2} \left[ \begin{array}{l} 0 \\ \sqrt{1-{z'}^2}\,\sqrt{1-y^2} \\ \sqrt{1-{z'}^2}\,y \\ z' \end{array}\right],
	\end{split}
\end{equation}
the kinematic function $K_0$ takes the form
\begin{equation}
	\begin{split}
		K_0 &= q^2\,k^2\,(1-\Omega^2) - Q^2\,\Big[ q^2\,(1-z^2) \\
		&  + k^2\,(1-{z'}^2) -2\sqrt{q^2}\sqrt{k^2}\,(\Omega-zz')\Big]\,,
	\end{split}
\end{equation}
where $\Omega = zz' + y\sqrt{1-z^2}\,\sqrt{1-{z'}^2}$
and the integral measure is
\begin{equation}
	\int_k = \frac{1}{(2\pi)^3}\,\frac{1}{2}\int_0^{\Lambda^2} dk^2\,k^2\,\int_{-1}^1 dz\sqrt{1-z^2} \int_{-1}^1 dy\,.
\end{equation}

After working out the projection~\eqref{gh-gl-projection},
the DSE takes the form
\begin{equation}
	\begin{split}
		a &= 1 + g^2 \frac{N_c}{2}\int_k a_1\,y\,\frac{k}{q}\sqrt{\frac{1-{z'}^2}{1-z^2}}\,\mI + (\dots)\,, \\
		b &= \frac{1}{2} + g^2\frac{N_c}{2}\int_k \left[ b_1 + a_1\,\frac{k}{Q}\,\frac{z'-z\Omega}{1-z^2}\right] \mI + (\dots)\,,
	\end{split}
\end{equation}
where $(\dots)$ denotes the contributions from the remaining loop diagrams.
Note that the integral for $a$ is suppressed by the factor $y$, which is integrated over a symmetric integral
and picks out the parts in $\mI$ that are odd in $y$; for the same reason the integrand is non-singular for $q\to 0$.
This is compatible with the observation that the loop contributions to the  ghost-gluon dressing are suppressed and $A$ is small.

Concerning the integral for $b$, one can employ the identity~\eqref{zy-womega} and write
\begin{equation}
	\frac{z'-z\Omega}{1-z^2} = w\,\sqrt{\frac{1-\Omega^2}{1-z^2}}\,.
\end{equation}
In the limit $Q^2\to 0$, the $z'$ dependence in $\mI$ drops out and the integration over $w/Q$ yields again a finite result.
Thus, the integrand is finite at $Q^2 = 0$, so the only possible singularities can come from $b$ itself
and must match on both sides of the equation. The equation for $b$ (and equivalently for $B$) thus becomes an inhomogeneous BSE,
\begin{equation}
	b =  [\dots] + g^2 \frac{N_c}{2}\int_k \mI\,b_1 \,,
\end{equation}
where the inhomogeneity $[\dots]$ is finite at $Q^2= 0$.
As discussed below Eq.~\eqref{bs-amp}, one can solve a homogeneous BSE for the Bethe-Salpeter amplitude $\varphi$,
which is the residue at the pole;
if that equation has a solution (i.e., if some eigenvalue becomes 1)  at a certain value of $Q^2$, then
there must be a corresponding pole in $b$.

We can solve the homogeneous BSE directly at $Q^\mu=0$, in which case the function $\mI$ becomes
\begin{equation}
	\mI = \frac{G(k^2)^2}{k^4}\,\frac{Z(l^2)}{l^4}\,q^2\,k^2\,(1-\Omega^2)\,.
\end{equation}
Since the ghost and gluon dressing functions therein were obtained from the Yang-Mills DSEs
with a tree-level ghost-gluon vertex, we set $a_2=1$ for consistency.
The resulting homogeneous BSE at $Q^2=0$ is
\begin{equation}\label{hom-bse-2}
	\varphi(q^2) = g^2 \frac{N_c}{2} \,q^2 \int_k \frac{G(k^2)^2}{k^2}\,\frac{Z(l^2)}{l^4}\,(1-\Omega^2)\,\varphi(k^2)\,.
\end{equation}
Note that this implies $\varphi(q^2\to 0) \propto q^2$, which is consistent with $b \propto q^2/Q^2$ being a dimensionless function.
For large $q^2$ we find $\varphi(q^2) \propto 1/q^2$ and the integral is convergent.
After employing the redefinitions leading to Eqs.~\eqref{dses-redef-2},
the BSE becomes identical to Eq.~\eqref{hom-bse} in the main text.

Returning to the question of a ghost-antighost symmetric ghost-gluon vertex,
in that case the most general form of the vertex is Eq.~\eqref{gh-gl-vertex-2} with $b = (q\cdot Q)\,b'$,
where the tree-level vertex corresponds to $a=1$ and $b=0$.
The tree-level contribution to the equation for $b$ therefore vanishes
and the integrand of the BSE~\eqref{hom-bse-2} picks up a factor $k\cdot Q/q\cdot Q = kz'/(qz)$,
which strongly suppresses the kernel. In particular, we find that the maximum value of the eigenvalue $\lambda_0$
for the scaling solution ($\alpha\to\infty$ in Fig.~\ref{fig-long-poles}) becomes $\lambda_0 \approx 0.13$,
so there is no pole at $Q^2=0$. Even if there was a pole,
the divergence of the function $b \propto q\cdot Q/Q^2$ would not be strong enough to
ensure the validity of Eq.~\eqref{PiT-condition}.
The emergence of longitudinal poles in the ghost-gluon vertex can thus be related to the
lack of the ghost-antighost symmetry.

\bibliographystyle{apsrev4-1-mod}
\bibliography{bib-ym}

\end{document}